\documentclass[useAMS,usenatbib]{mn2e}

\usepackage{graphicx}
\usepackage{amstext}
\usepackage{url}

\def   \araa {{\rm {ARA\&A}}}
\def   \apj {{\rm {ApJ}}}
\def   \icarus {{\rm {Icarus}}}
\def   \apjs {{\rm {ApJS}}}
\def   \apss {{\rm {Ap\&SS}}}
\def   \aap {{\rm {A\&A}}}

\def   \aaps {{\rm {A\&AS}}}

\def   \mnras {{\rm {MNRAS}}}

\title[Chemistry in a grav. unstable
protoplanetary disc]{Chemistry in a gravitationally unstable protoplanetary disc}

\author[J.~D.~Ilee et al.]
{\parbox{\textwidth}{J.~D.~Ilee$^{1}$\thanks{E-mail: \texttt{pyjdi@leeds.ac.uk}},
A.~C.~Boley$^{2}$,
P.~Caselli$^{1}$,
R.~H.~Durisen$^{3}$,  
T.~W.~Hartquist$^{1}$ and 
J.~M.~C.~Rawlings$^{4}$}\vspace{0.4cm}\\
\parbox{\textwidth}{
$^{1}$School Of Physics \& Astronomy, University Of Leeds, Leeds LS2 9JT, UK\\
$^{2}$Department of Astronomy, University of Florida, 211 Bryant Space Science Center, PO Box 112055, USA\\
$^{3}$Department of Astronomy, Indiana University, 727 East 3rd Street, Swain West 319, Bloomington, IN 47405, USA\\
$^{4}$Department of Physics \& Astronomy, University College London, London WC1E 6BT, UK
}}

\begin{document}

\date{Accepted 2011 July 15.  Received 2011 July 4; in original form 2011 May 1}

\pagerange{\pageref{firstpage}--\pageref{lastpage}} \pubyear{0000}

\maketitle

\label{firstpage}

\begin{abstract}

Until now, axisymmetric, $\alpha$-disc models have been adopted for
calculations of the chemical composition of protoplanetary
discs. While this approach is reasonable for many discs, it is not
appropriate when self-gravity is important. In this case, spiral waves
and shocks cause temperature and density variations that affect the
chemistry. We have adopted a dynamical model of a solar-mass star
surrounded by a massive (0.39 M$_{\odot}$), self-gravitating disc,
similar to those that may be found around Class 0 and early Class I
protostars, in a study of disc chemistry.  We find that for each of a
number of species, e.g.\  H$_2$O, adsorption and desorption dominate the
changes in the gas-phase fractional abundance; because the desorption
rates are very sensitive to temperature, maps of the emissions from
such species should reveal the locations of shocks of varying
strengths. The gas-phase fractional abundances of some other species,
e.g.\ CS, are also affected by gas-phase reactions, particularly in
warm shocked regions. We conclude that the dynamics of massive discs
have a strong impact on how they appear when imaged in the emission
lines of various molecular species.

\end{abstract}


\section[Introduction]{Introduction}

In the last several years, the Plateau de Bure Interferometer
\citep{dutrey_chemistry_2007, henning_chemistry_2010} and the
Submillimeter Array \citep{qi_resolving_2008} have
provided images of protoplanetary discs in an increasing variety of
molecular line emissions. The field will be advanced further by the
completion of the Submillimeter Array Disc Imaging Survey
\citep{oberg_disk_2010}. It will be revolutionised through the use of
the Atacama Large Millimetre/submillimetre Array (ALMA), which will
allow much weaker lines to be investigated, and thus will enable the
characterisation of a more diverse range of processes
\citep{guilloteau_new_2008, semenov_chemical_2008}.

Many authors \citep[e.g.,][]{aikawa_evolution_1999, aikawa_warm_2002,
  markwick_molecular_2002, bergin_effects_2003, gorti_models_2004,
  ilgner_transport_2004, nomura_molecular_2005,
  willacy_turbulence_2006, willacy_chemistry_2007,
  nomura_effects_2009, woitke_hot_2009, walsh_chemical_2010,
  fogel_chemistry_2011, heinzeller_chemical_2011} have presented
  computational results for the chemical composition of protoplanetary
  discs. Because $\alpha$-disc models were adopted, these studies are
  relevant to protoplanetary discs that transport mass locally, both
  instantaneously and averaged in time, and by extension, to discs
  that possess global axisymmetry.  For this reason,
  $\alpha$-disc models may be inappropriate during early phases of gas
  accretion on to protoplanetary discs if discs become massive enough
  to trigger gas phase gravitational instabilities (GIs)
  \citep{vorobyov_origin_2005, vorobyov_burst_2006,
  vorobyov_secular_2009, boley_two_2009}. GIs produce dynamic
  nonaxisymmetric structure in the form of spiral waves leading to
  shocks \citep[see][for a review]{durisen_gravitational_2007} and
  possibly, under some conditions, to bound protoplanetary fragments
  \citep[][]{boss_gas_2001, mayer_evolution_2004,
  mayer_fragmentation_2007, stamatellos_brown_2007, boley_clumps_2010,
  vorobyov_formation_2010, boley_on_2010}. Fragmentation conditions
  are still under debate \citep[see, e.g.,][]{boss_testing_2007,
  cai_giant_2010,podolak_evolution_2010, meru_non_2011,
  lodato_resolution_2011} but, for non-fragmenting cases, it is
  generally agreed that gravitational torques caused by spiral waves
  lead to significant mass transport \citep[][]{gammie_nonlinear_2001,
  lodato_testing_2004, mejia_thermal_2005, boley_thermal_2006,
  boley_three_2007, cossins_characterising_2009,
  michael_submitted_2011}.

Such mass transport may have implications for disc driven outbursts
and recent, successful models for FU Orionis outbursts have been
developed for the assumption that GIs dominate mass transport outside
disc radii of a few AU in non-fragmenting discs
\citep{armitage_episodic_2001, zhu_nonsteady_2009,
zhu_longI_2010}. Simulations \citep{zhu_longII_2010} including layered
accretion in the Dead Zone \citep{gammie_layered_1996} as well as
transport by GIs show that, when discs accrete from rapidly rotating
cloud cores, disc masses can become comparable to those of their
central stars. Simple $\alpha$-disc models for disc evolution overlook
the episodic heating induced by GI spiral shocks in massive discs
\citep{boley_gravitational_2008}. These shocks can cause desorption of
volatiles from dust grains and can trigger gas-phase chemical
reactions that would not otherwise occur. Both effects can produce
observable chemical signatures of disc dynamics.

In this paper, we present results from a chemical model of a massive,
young protoplanetary disc in which GIs cause the formation of spiral
waves. Section \ref{sec:methods} outlines the physical and chemical
models we have utilised, and the assumed initial conditions.  Section
\ref{sec:results} contains results for the time-dependent fractional
abundances within an individual fluid element and column density maps
of the entire disc for different species. A comparison of results from
this model with those of other models of disc chemistry are also
given.  Finally, Section \ref{sec:conclusions} presents conclusions
based on the results and comments on further avenues of research.

\section[Methods]{Methods}
\label{sec:methods}

\subsection{The dynamical model}

We use a hydrodynamic simulation of a massive ($0.39\,$M$_{\sun}$)
protoplanetary disc as the basis for the physical input to our
chemical model. Most of the mass in the disc initially extends from a
radial distance $r\sim7$ to $50\,$au from the central, solar-mass
protostar.  The system represents an early Class I object, which would
likely evolve into an F-type main sequence star.  The disc is modelled
with \textsc{chymera} \citep{boley_phd_2007}, which solves the
equations of hydrodynamics with self-gravity on a fixed, cylindrical
Eulerian grid.  The computational domain has 256, 128 and 64 zones in
$r$, $\phi$ and $z$, respectively, with a physical resolution of
$\Delta r=\Delta z=0.25\,$au.  Mirror symmetry about the midplane is
assumed.  All grid boundaries have outflow conditions allowing mass to
leave the global disc simulation, where the inner boundary is set to
approximately $4\,$au, but no mass is allowed to enter. The central
star is allowed to move freely, and the equation of motion for it is
integrated as described in \citet{boley_two_2009}.  We use the
\citet{boley_internal_2007} equation of state with a fixed
ortho-to-para ratio of hydrogen molecules of 3:1 (which is only
relevant for the evolution of the physical model).  The fractional
mass abundances of hydrogen, helium and more massive elements are set
to $X=0.73$, $Y=0.25$, and $Z=0.02$, respectively. Because the
hydrogen is almost entirely in H$_2$, the mean molecular weight is
$2.33\,$amu.  For cooling, we use the radiative cooling approximation
described in \citet{boley_two_2009}.  An incident radiation field on
the disc is assumed to have a black body spectrum with a temperature
varying as $T_{\rm irr}= 140 (r/{\rm au})^{-0.5} + 10\,$K to account
for heating by the star.

\begin{figure}
\centering
\includegraphics[width=\columnwidth]{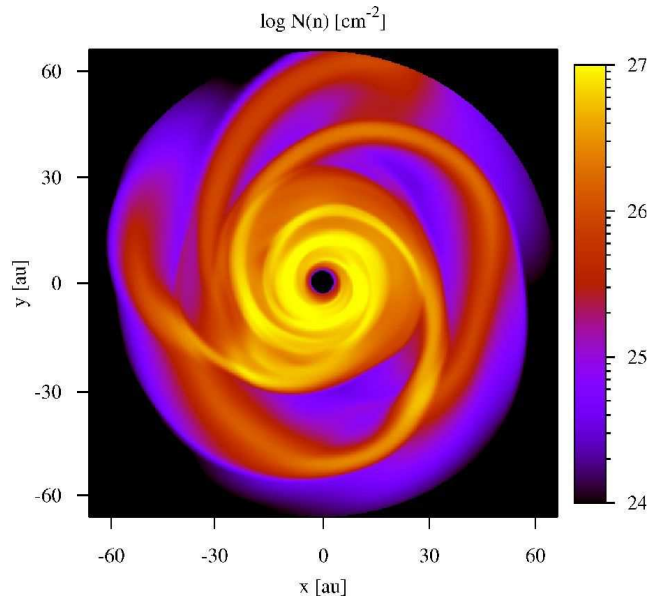}
\caption{Column density of nuclei (see Eqn. \ref{eqn:n} for the relationship
between number density and mass density) in the
disc viewed from above at the end of the simulation, $t = 388\,$yr.}
\label{fig:Nnden}
\end{figure}

The initial disc model is prepared using the method described in
\citet{boley_gravitational_2008}.  An analytic disc model is first
created, with a \citet{toomre_on_1964} $Q\sim1$ for most of the disc.
The spiral instability sets in around $Q\sim1.7$
\citep{durisen_gravitational_2007}, so the initial model represents
conditions before a strong burst of gravitational instabilities, as
might be expected during the earliest stages of disc evolution.  A
flat $Q$ profile with the irradiation law noted above requires the
surface density to follow, roughly, $\Sigma\propto r^{-1.75}$.  This
initial configuration is assumed to undergo Keplerian rotation and
have an adiabatic index of 5/3. These assumptions about the initial
rotation and value of the adiabatic index can lead to large radial
oscillations at the start of the simulation due to the associated
deviations from equilibrium.  To avoid some of this behavior, we
evolve the analytic model in \textsc{chymera} at low azimuthal
resolution (8 zones) for 500$\,$yr, which is about two orbital periods
near $r\sim45\,$au.  When the disc is loaded on to the higher
resolution grid, a 5\% cell-to-cell random noise is added to the
density distribution, which seeds the growth of nonaxisymmetric
structure.  This is considered to be the time $t=0$ in the initial
disc.  The disc is then run at higher azimuthal resolution for an
additional $388\,$yr. Figure \ref{fig:Nnden} shows the column density
of nuclei for the disc viewed from above at the end of the simulation.

\begin{figure}
\centering
\includegraphics[width=0.85\columnwidth]{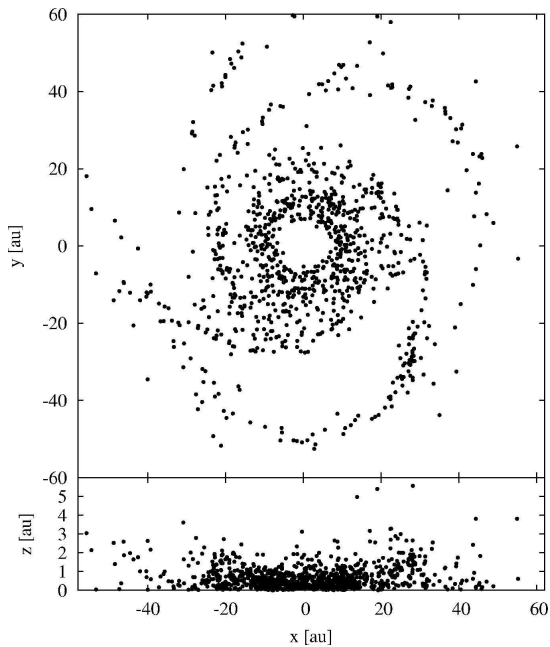}
\caption{The final location of the fluid elements used to sample the disc, at $t=388\,$yr.}
\label{fig:discparcels}
\end{figure}

At the start of the full simulation ($t=0$), 1000 fluid elements are
randomly distributed throughout the disc, weighted by mass.  These
fluid elements are evolved along with the main simulation, and their
thermal histories are recorded, as described in
\citet{boley_gravitational_2008}. The division of the disc
into more fluid element would have allowed the capture of rare events
such as shocks that are atypically strong for the radii at which they
occur.  This would lead to slightly greater average gas phase
abundances of species that are hard to desorb. However, this would not
affect our overall conclusions.  By the end of the simulation, 5
fluid elements were lost through grid boundaries, leaving 995 elements
with complete thermal histories. 963 of these were used to study the
chemical evolution in the disc.  Results for 16 of the 32
remaining fluid elements were not used to generate chemical results
because the jump, for each of those 16, in the physical conditions
from those at the end of the dynamical calculation back to those at
the start of the dynamical calculation created difficulties for the
integration of the chemical rate equations. Results for the other 16
were not used because spiked extrema in them caused problems for the
chemical integration. These extrema occurred at temperatures of around
$30\,$K - $40\,$K. The 32 elements were randomly distributed at the
end of the simulation, and had not passed through shocks near the end
of the dynamical calculation, so their influence on the chemical
results was negligible.  Figure \ref{fig:discparcels} shows the
distribution of the final locations of the fluid elements used in the
chemical modelling.

\begin{figure}
\centering
\includegraphics[width=0.85\columnwidth]{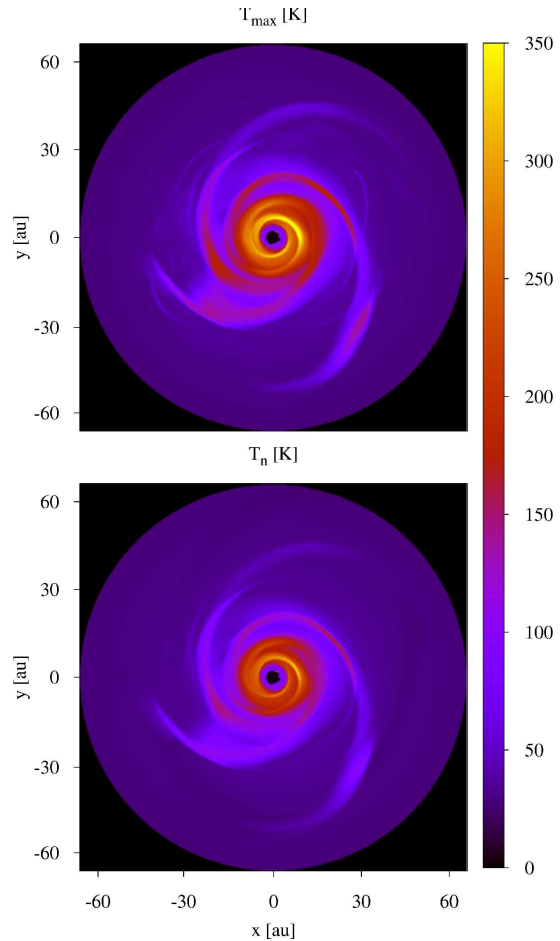}
\caption{Maximum line of sight temperature (top) and mass averaged
temperature (bottom) of the disc at $t=388\,$yr.}
\label{fig:temps}
\end{figure}

Figure \ref{fig:temps} shows the maximum line-of-sight temperature and
the mass-averaged temperature $T_{n} = \left( \int n T \, \rm{d}z
\right) / \left( \int n \, \rm{d}z \right)$ along the line-of-sight
for a face-on view of the final disc. Here $n$ is the number density
of nuclei (see Eqn. \ref{eqn:n}), $T$ is the temperature and d$z$ is
an infinitesimal path length along the line-of-sight. The maximum and
mass-weighted line-of-sight temperatures were calculated from the full
hydrodynamic model results, rather than from the properties of the
massive fluid elements used in the chemical calculations, because the
full hydrodynamic model results provide higher spatial resolution.

The spiral structure is clearly seen in the maximum line-of-sight
temperature map. The structure is also seen in the mass-averaged
temperature map, but is somewhat less sharply defined.  The difference
in the maps is due to the high temperature region being somewhat, but
not highly, limited in extent.  It lies near the midplane and contains
only a fraction of the disc mass.

\begin{figure*}

\centering
\includegraphics[width=0.9\columnwidth]{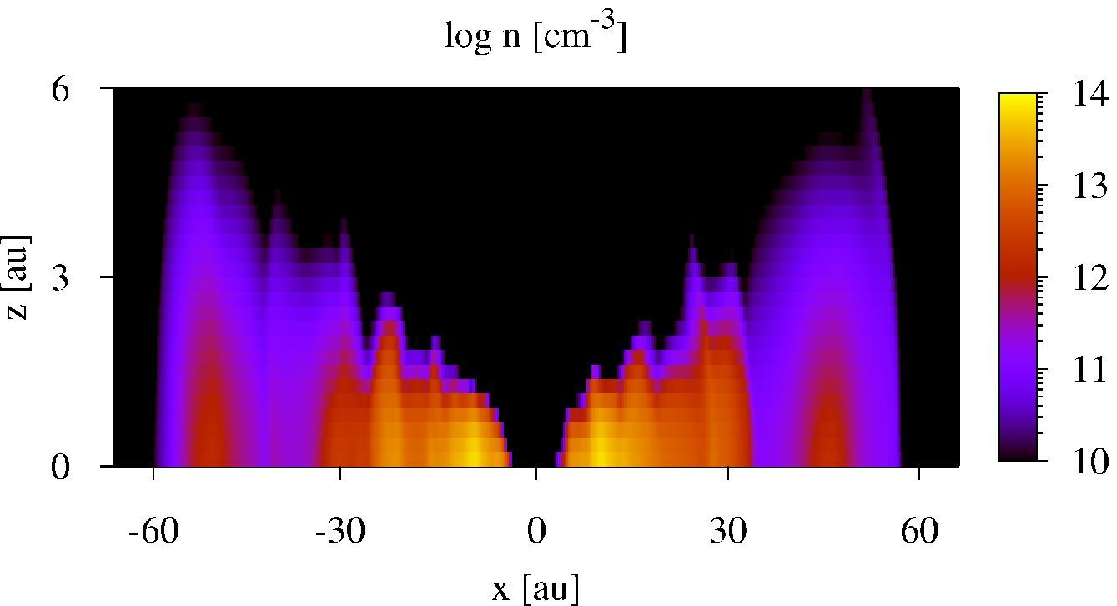}
\hspace{0.2in}
\includegraphics[width=0.9\columnwidth]{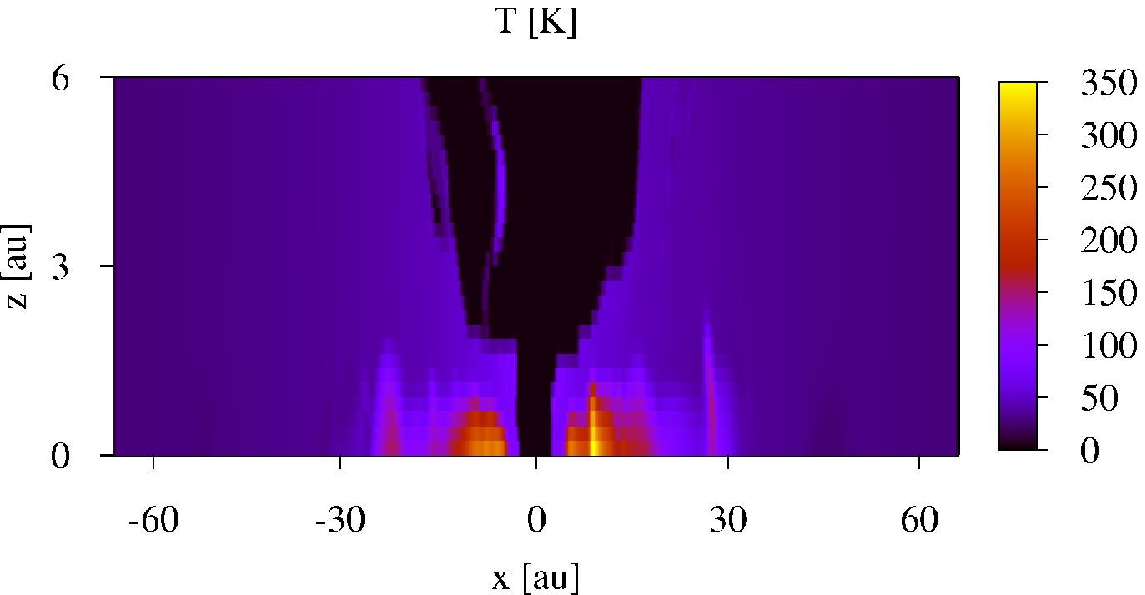}

\vspace{0.2in}

\includegraphics[width=0.9\columnwidth]{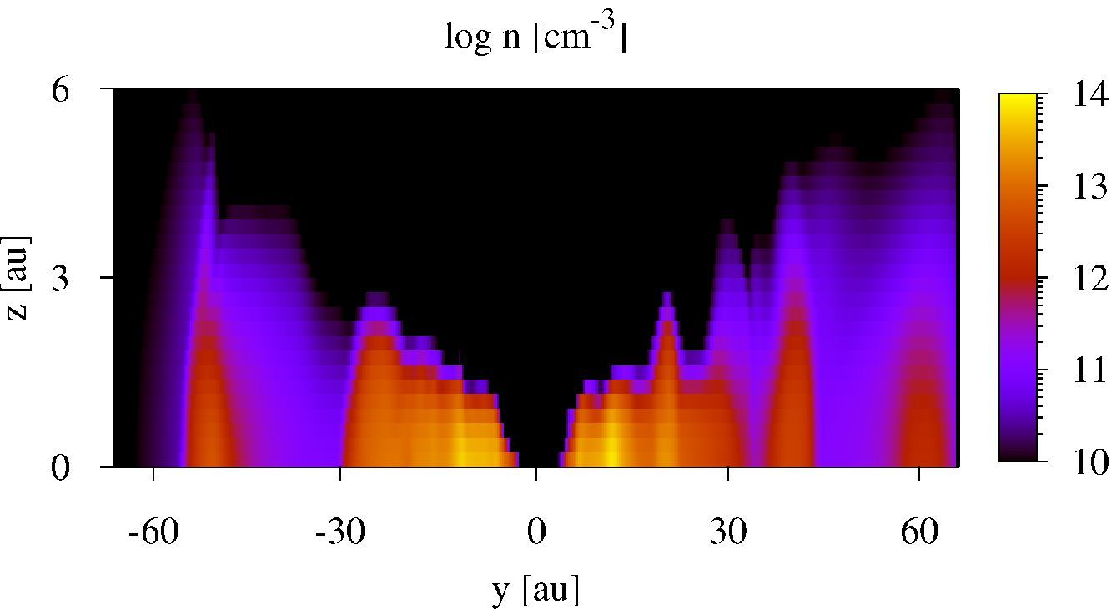}
\hspace{0.2in}
\includegraphics[width=0.9\columnwidth]{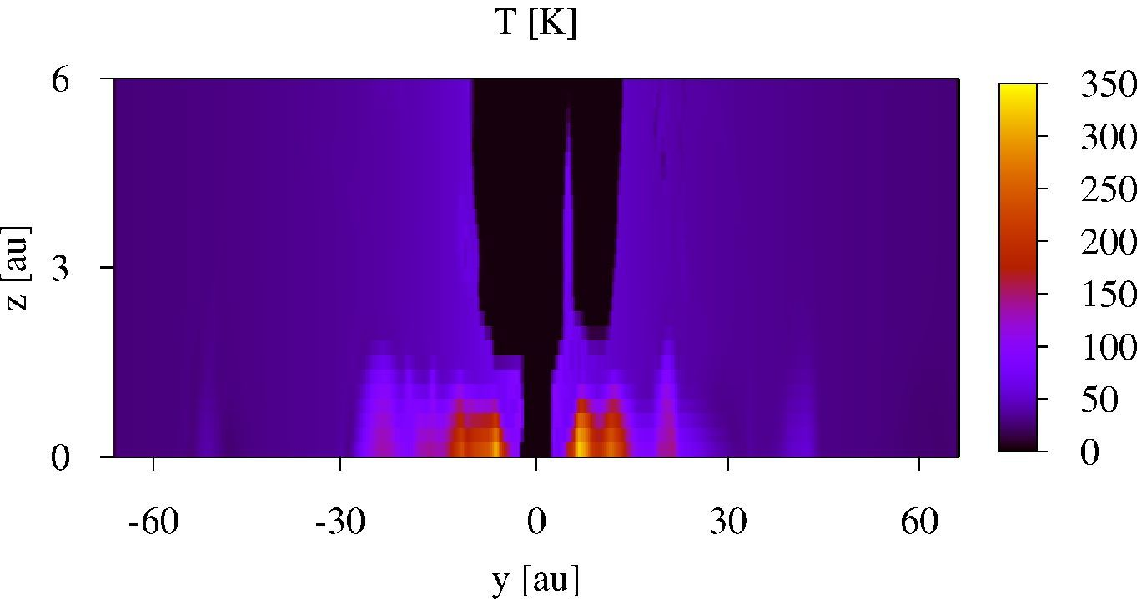}
\caption{Slices of the disc interior showing number densities and
  temperatures at $t=388\,$yr.  The $x$-axis here corresponds to a
  slice along the $x$-axis of Fig. \ref{fig:Nnden} where $y=0$, and
  vice versa.}
\label{fig:tnslice}
\end{figure*}

Figure \ref{fig:tnslice} shows results, taken from the complete
hydrodynamical data, for the temperature and number density of nuclei
in planes containing the rotation axis and either the North-South axis
($y$) or the East-West axis ($x$) from Fig. \ref{fig:discparcels}.
The temperature is less than $150\,$K outside the inner $20\,$au of
the disc, while within it, the temperature can reach up to $350\,$K.
The number density plot shows that the highest density gas lies in a
ring-like structure at approximately $10\,$au from the centre.

\begin{figure}
 \centering
\includegraphics[width=\columnwidth]{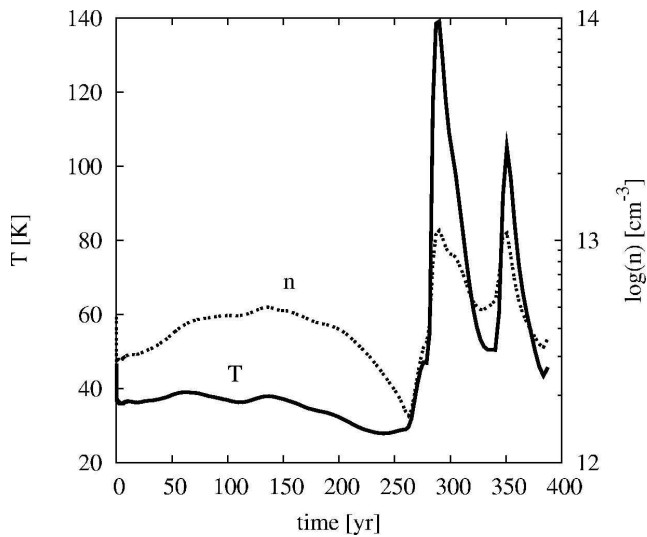}
  \caption{Temperature and number density history of a fluid element
  from the disc.  This particular element encounters a shock at about
  $270$ and $350\,$yr.}
\label{fig:tracks}
\end{figure}

The temperature and density histories of the fluid elements shown in
Fig. \ref{fig:discparcels} form the physical input to the chemical
model.  Figure \ref{fig:tracks} shows the evolution of the temperature
and density of one of the fluid elements.  This element begins at
$0.8\,$au above the midplane, approximately $30\,$au from the centre
of the disc.  It follows a nearly circular path that slowly increases in
radius.  After one-and-a-half orbits, or $270\,$yr, it encounters the
first shock.  Both shocks do not affect the circular motion, but
instead cause the element to rise to approximately $1.8\,$au above the
midplane.
The results for this track provide a useful example, because the fluid
element experiences a relatively quiescent period for $270\,$yr,
before encountering the shocks.  These shocks rapidly raise the
temperature to a maximum of $140\,$K and the number density to
$10^{13}\,$cm$^{-3}$. The temperature and density appear to
peak at the same time and drop together. This is due to the increase
in the pressure caused by a shock driving an expansion of gas parallel
to the shock and obliquely to the mid-plane of the disc that begins
almost immediately behind the shock \citep[see][]{boley_gravitational_2008}.

\subsection{The chemical model}

The chemical computation is based on a network of rate equations
involving 1334 reactions and 125 species containing the elements H,
He, C, O, N, Na and S.  These reactions were originally selected from
a subset of the UMIST Rate 95 database \citep{millar_umist_1997}. Data
from the Kinetic Database For Astrochemistry,
KIDA\footnote{\url{http://kida.obs.u-bordeaux1.fr}}, were used to
update some of the rates and rate coefficients.  Thermal desorption
processes (discussed in detail below) and some three-body reactions
(see Appendix \ref{sec:3br}) were added to the network.

At each time-step in the chemical network, the temperature and mass
density of each fluid element are derived with cubic spline fits to
the data provided for the element from the physical model. These
interpolated values of the temperature and mass density are then used
in the time-integration of the rate equations. The number density of
nuclei, $n$, is calculated from the mass density, $\rho$, by
\begin{equation}
\label{eqn:n}
n = \frac{N_{\rm{A}}}{M} \rho = n_{\rm{H}} + n_{\rm{He}} + n_{\rm{Z}},
\end{equation}
where $n_{\text{H}}$, $n_{\text{He}}$ and $n_{\text{Z}}$ are the
number densities of hydrogen, helium and more massive nuclei,
respectively.  We have assumed a molar mass of $M = 1.28$ g
mol$^{-1}$, which is appropriate for a relative abundance of
helium-to-hydrogen of 0.09 with trace amounts of more massive
elements. Integrations were performed with the \textsc{dvode} package
\citep{brown_vode_1989} to obtain the fractional abundance,
$X(i)=n(i)/n$, of each species.

The rate equation for the gas-phase fractional abundance of the $i$th
species is
\begin{eqnarray}
\frac{\rm{d}}{\rm{d}t} X(i) = \sum_{j,l,m} k(j) X(l) X(m) n - \sum_{j',m} k(j') X(i) X(m) n \\
\nonumber - 2 \sum_{j''} k(j'') X(i)^{2} n - \Gamma_{\rm{cr}}(i)X(i)
+ S_3(i) + S_{\rm{a,d}}(i),
\end{eqnarray}
where $k(j)$ is the rate coefficient of the $j$th reaction and the
summations are restricted so that only reactions involved in the
formation or removal of the $i$th species are included.
$\Gamma_{\rm{cr}}(i)$ is the rate at which direct cosmic-ray
ionisation and cosmic-ray-induced photoemission remove species $i$.
$S_3(i)$ and $S_{\rm{a,d}}(i)$ are source terms due to three-body
reactions and adsorption on to and desorption from grain surfaces,
respectively.  We assume that the only third body of importance is
H$_2$, an assumption which introduces only an insignificant error.
Thus,
\begin{eqnarray}
S_3(i) = \sum_{j,l,m} k(j) X(l) X(m) X(\rm{H}_{2}) n^{2} \\
\nonumber - \sum_{j',m} k(j') X(i) X(m) X(\rm{H}_{2}) n^{2} \\
\nonumber - 2 \sum_{j''} k(j'') X(i)^{2} X(\rm{H}_{2}) n^{2}.
\end{eqnarray}

\subsubsection{Rate coefficients and rates}

Here we present a brief summary of the forms of the gas-phase rate
coefficients and rates that we have adopted. More detail can be found
in the primary paper on the UMIST Rate 95 database
\citep{millar_umist_1997}. The reaction rates and coefficients
that we use are provided
online\footnote{\url{http://www.ast.leeds.ac.uk/~pyjdi/chemdisc/network.txt}}.

The rate coefficient for the $j$th two- or three-body reaction is of
the standard Arrhenius form
\begin{equation}
k(j)=\alpha(j) \left(\frac{T}{300}\right)^{\beta(j)}\exp{\left(\frac{-\gamma(j)}{T}\right)},
\end{equation}
where $\alpha(j)$ is the room temperature rate coefficient of the
reaction (at $300\,$K), $\beta(j)$ describes the temperature
dependence and $\gamma(j)$ is the activation energy of the reaction.

The rate for the destruction of species $i$ due to cosmic rays is
given by
\begin{equation}
\Gamma_{\rm{cr}}(i) = a(i) \zeta + \frac{\zeta P(i)}{1-A},
\end{equation}
where $a(i)$ is a proportionality constant, $\zeta = 10^{-17}\,\rm{
s}^{-1}$ is the cosmic ray ionisation rate, $P(i)$ is a constant and
$A$ is the dust grain albedo, which we take to be 0.5. Because
the two terms on the right represent the physically distinct processes
of direct cosmic ray induced ionisation and of destruction due to
photoemission induced by the collisions of molecular hydrogen with
energetic electrons produced by the ionisation, data tables give
$a(i)$ and $P(i)$ separately.

Though photoabsorption of radiation from external sources will affect
the chemistry in the outer layers of a disc, we focus on the bulk of
the disc material and assume that it is well shielded from external
sources of photons. Thus, we neglect photoabsorption, other than that
of cosmic-ray-induced photoemission.

\subsubsection{Gas-grain reactions}

With minor exceptions, we assume that no surface chemistry occurs,
though we recognise that surface chemistry is important in
establishing the chemical initial conditions that we adopt. The
exceptions concern the neutralisation of ions, which we assume occurs
when they are adsorbed.  While studies of the chemistry of discs
including surface reactions and grain evolution have been performed
\citep{schreyer_chemistry_2008,henning_chemistry_2010,semenov_chemistry_2010,
vasyunin_impact_2011,semenov_chemical_2011}, such chemistry remains
very uncertain.  This is why we have assumed that the initial
abundances reflect those of cometary ices (see Section
\ref{sec:initial}).

$S_{\rm{a,d}}(i)$ has contributions due to adsorption and desorption,
and we take
\begin{equation}
S_{\rm{a,d}}(i) = S_{\rm{d}}(i) - S_{\rm{a}}(i),
\end{equation}
where
\begin{equation}
S_{\rm{a}} = \pi a^2 S(i) \eta \sqrt{\frac{8kT}{\pi \mu m_{\rm{H}}}}
X(i) n.
\label{eqn:ads}
\end{equation}
The sticking coefficient $S(i)$ is set to unity, and we take the ratio
of the number density of dust grains to the number density of nuclei
to be $\eta = 3.3\times10^{-12}$, which is appropriate for a dust
grain radius and grain-to-gas mass ratio of about $a=0.1$ $\mu \rm{m}$
and 0.01, respectively.  $T_{\rm{g}}$ is the gas temperature, $k$ is
the Boltzmann constant, $\mu$ is the molecular mass in amu and
$m_{\rm{H}}$ is the mass of a hydrogen atom.

We treat thermal desorption in the same way as
\cite{visser_chemical_2009}, and
\begin{equation}
S_d(i) = 1.26\times10^{-21} (\sigma / \text{cm}^{-2}) f(i) \nu_{0}(i)
\; \exp \left( \frac{-E_{\rm{b}}(i)}{k\:T_{\rm{d}}} \right),
\label{eqn:des}
\end{equation}
where $\sigma$ is surface density of binding sites, which we take to
be $1.5\times10^{15}\,$cm$^{-2}$. $E_{\rm{b}}(i)$ is the binding
energy of the $i$th species on the surface of the dust grain,
$T_{\rm{d}}$ is the dust temperature (which we assume to be in
equilibrium with $T_{\rm{g}}$), the factor $f(i)$ represents the
fraction of the surface of the dust grain covered by the $i$th
species, given by
\begin{equation}
f(i) = \text{min} \left( 1, \frac{X^{\rm{s}}(i)}{\eta \:N_{\rm{b}}} \right),
\end{equation} 
where $X^{\rm{s}}(i)$ is the solid fractional abundance of the $i$th
species and $N_{\rm{b}}$ is the typical number of binding sites per
grain, taken to be $10^{6}$.  The characteristic vibrational frequency
of the species attached to the grain is given by
\begin{equation}
\nu_{0}(i) = \sqrt{\frac{2 \sigma E_{\rm{b}}(i)}{m(i) \pi^{2}}},
\end{equation}
where $m(i)$ is the mass of the $i$th species
\citep{hasegawa_models_1992}.  The binding energies were taken from
\citet{hollenbach_Water_2009}, references therein, and the OSU
database\footnote{\url{http://www.physics.ohio-state.edu/~eric/research.html
}}.

 \subsection{Chemical initial conditions}
\label{sec:initial}
The initial gas-phase fractional abundances were assumed to have
the same ratios to one another as the fractional abundances of the
corresponding ices in comet Hale-Bopp as given by
\citet{ehrenfreund_organic_2000}, and they are given in Table
\ref{tab:ics}. Because comets are thought to have undergone
significant chemical processing at the edges of discs, this assumption
may not be entirely appropriate. However, comparisons between the
compositions of cometary ices and interstellar ices imply a general
consistency between the two, though some discrepancies exist
\citep{ehrenfreund_organic_2000,ehrenfreund_iso_2000}.

\begin{table}
\caption{Initial fractional abundances ($X(i) =
n(i) / n_{\text{H}}$). Note that $a(b) \equiv a\times
10^{b}$.}
\label{tab:ics}
\begin{minipage}[t]{0.45\linewidth}
\centering
\begin{tabular}{c|c}
\hline
Species & Abundance \\
\hline
He	&       1.00(-1) \\
C     &       3.75(-4) \\
CO   &     3.66(-5) \\
CH$_{4}$  &     1.10(-6) \\
N      &   1.15(-4) \\
NH$_{3}$   &   3.30(-6) \\
O      &  6.74(-4) \\
H$_{2}$O  &    1.83(-4) \\
Na    &   3.50(-5) \\
\hline
\end{tabular}
\end{minipage}
\hspace{0.1cm}
\begin{minipage}[t]{0.45\linewidth}
\centering
\begin{tabular}{c|c}
\hline
Species & Abundance \\
\hline
H$_{2}$CO  &   1.83(-6) \\
CO$_{2}$    &  3.67(-5) \\
HCN    &  4.59(-7) \\
HNC   &   7.34(-8) \\
S   &     1.62(-5) \\
H$_{2}$S    &  2.75(-6) \\
SO   &    1.47(-6) \\
SO$_{2}$   &   1.84(-7) \\
OCS   &   3.30(-6) \\
\hline
\end{tabular}
\end{minipage}
\end{table}

\subsection{Timescales}

As mentioned above, the hydrodynamical simulation following the disc
evolution is run at higher resolution for
$388\,$yr. This is far longer than the orbital period
of about $4\,$yr at the inner boundary and somewhat shorter than the
orbital period of $390\,$yr at $60\,$au.  Nearly all of the mass is at less than
$50\,$au where the period is about $300\,$yr, and the gravitational
instability develops fully and leads to well established spiral
structure in about $200\,$yr. The spiral structure roughly co-rotates
with the material between about $30$ and $40\,$au; the orbital period
at $35\,$au is $180\,$yr. The disc is not steady, and the mass infall rate
varies with radius and time from a factor of a few smaller than to
a factor of a few larger than 10$^{-4}\,$M$_{\sun}$, which implies
a radial flow time of the order of $10^{4}\,$yr. These timescales are
much longer than
the shortest chemical timescales which are associated with adsorption
and desorption; using Eqn. 7, one finds that the adsorption timescale
for a number density of 10$^{13}\,$cm$^{-3}$ is roughly an hour. The
desorption timescale is very much larger or smaller, depending on the
temperature.

$388\,$yr is sufficient to allow the spiral
shocks to develop fully.  This disc is already exhibiting spiral
structure that is typical in GI-active discs over many orbits.  The
purpose of the hydrodynamical simulation is to provide self-consistent
shock profiles for the chemical model, which was achieved in a
relatively short period of evolution.

We follow the chemistry of each fluid element through ten cycles of
the temperature and number density history associated with it (e.g.,
the history shown in Fig. \ref{fig:tracks}), where in each subsequent
cycle, we used the final fractional abundances from the previous cycle
as initial input.  Of course, this resulted in rapid variations in the
physical conditions, as each new cycle began. However, we followed the
chemical evolution in this way to keep the amount of data produced by
the model at manageable levels. We found that the chemical
behaviour for a fluid element became periodic by the time that the
chemistry had gone through ten cycles. All results presented here are
from the final cycle, and $t=0$ occurs at the beginning of that cycle.

At the start of each integration, we used a logarithmically increasing
time-step, which was initially $90\,$s. This allowed initially rapid
reactions, such as the adsorption of species on to grains at the
beginning of the first cycle, to be followed with sufficient
resolution. We limited the maximum time-step to approximately
$10^{3}\,$s to avoid missing details in rapidly changing shock
features.

\section[Results]{Results}
\label{sec:results}

The main results of the paper are column density maps for a variety of
gas-phase species. To obtain them, we followed the chemical evolution
of each fluid element. Before presenting the column density maps, we
consider the chemistry in one fluid element as an illustrative example
of the effect of shocks on the chemistry of the material.

\subsection{An individual fluid element}
\label{sec:individual}

\begin{figure*}
 \centering
\includegraphics[width=0.8\columnwidth]{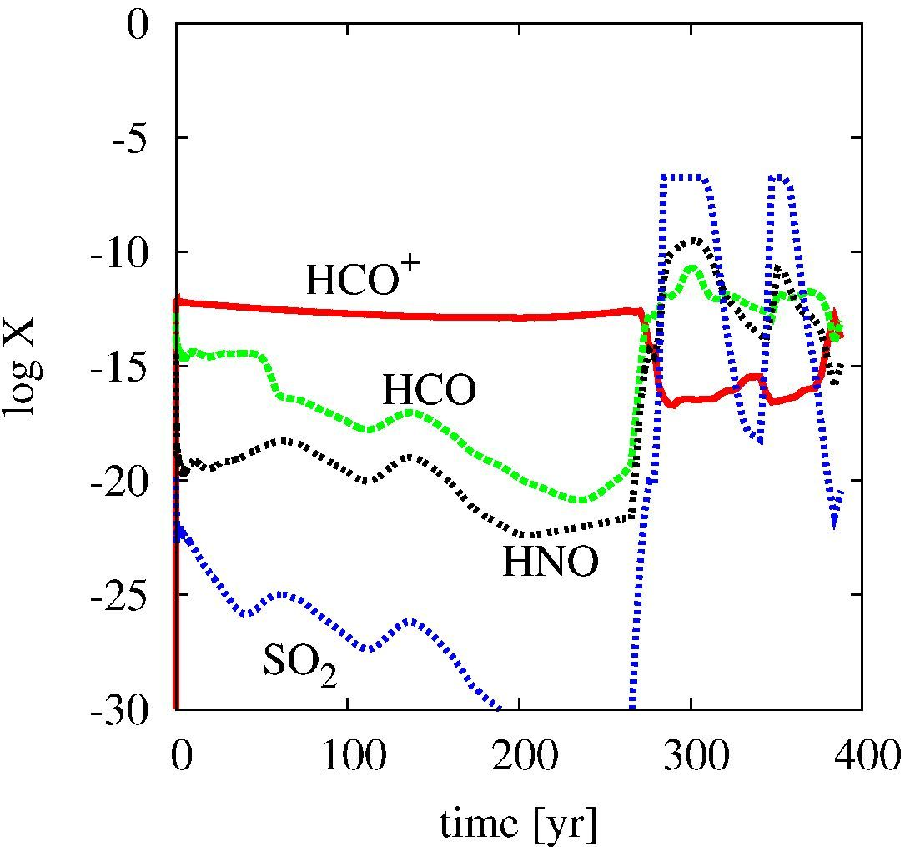}
\hspace{0.2in}
\includegraphics[width=0.8\columnwidth]{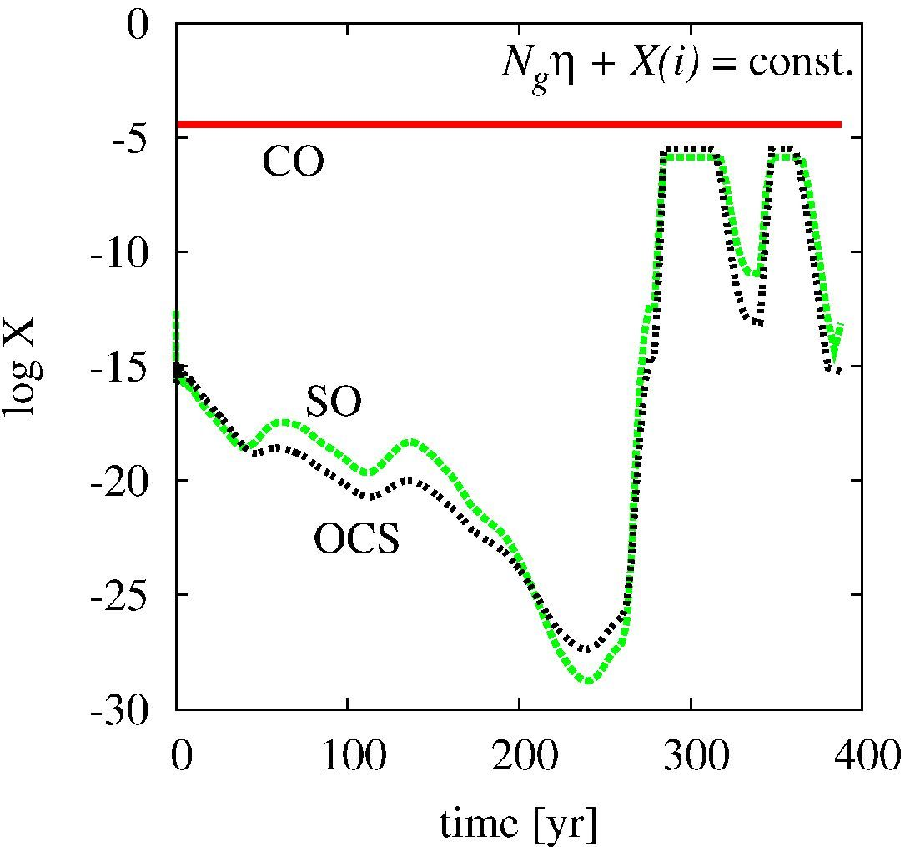}

\vspace{0.2in}

\includegraphics[width=0.8\columnwidth]{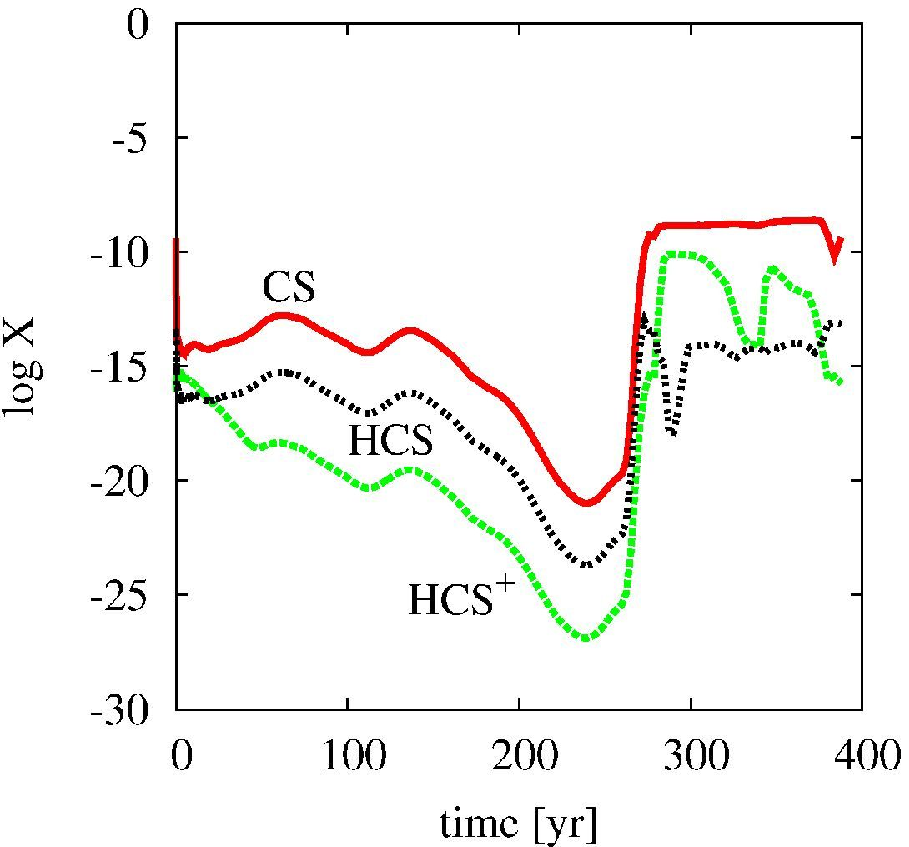}
\hspace{0.2in}
\includegraphics[width=0.8\columnwidth]{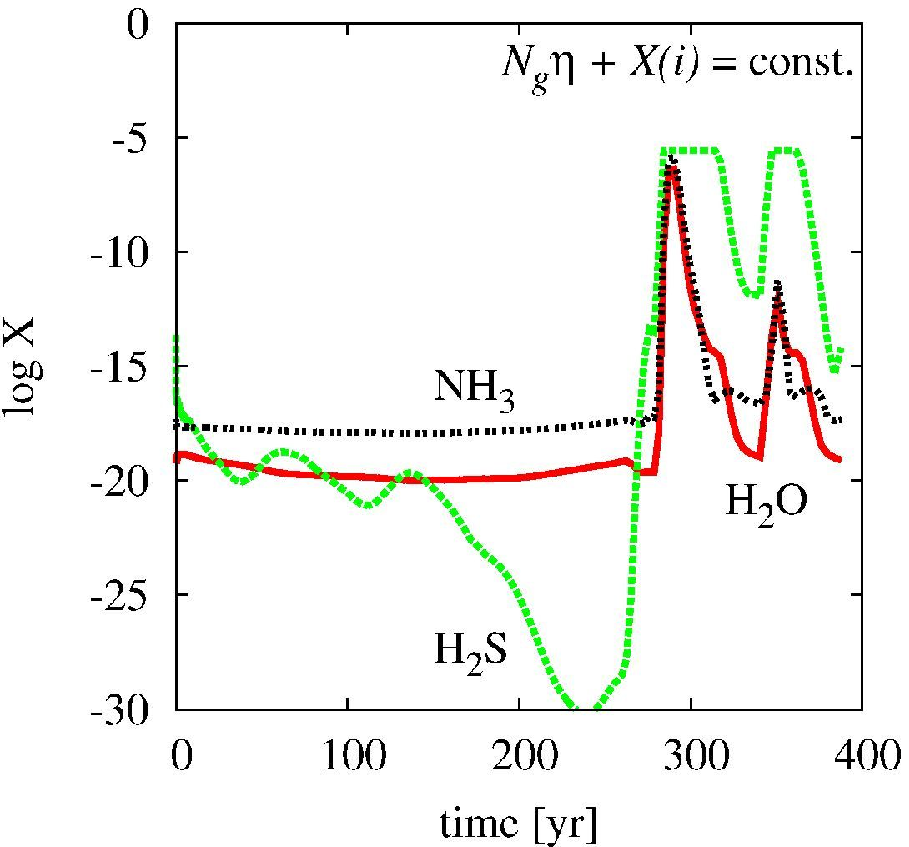}

\vspace{0.2in}

\includegraphics[width=0.8\columnwidth]{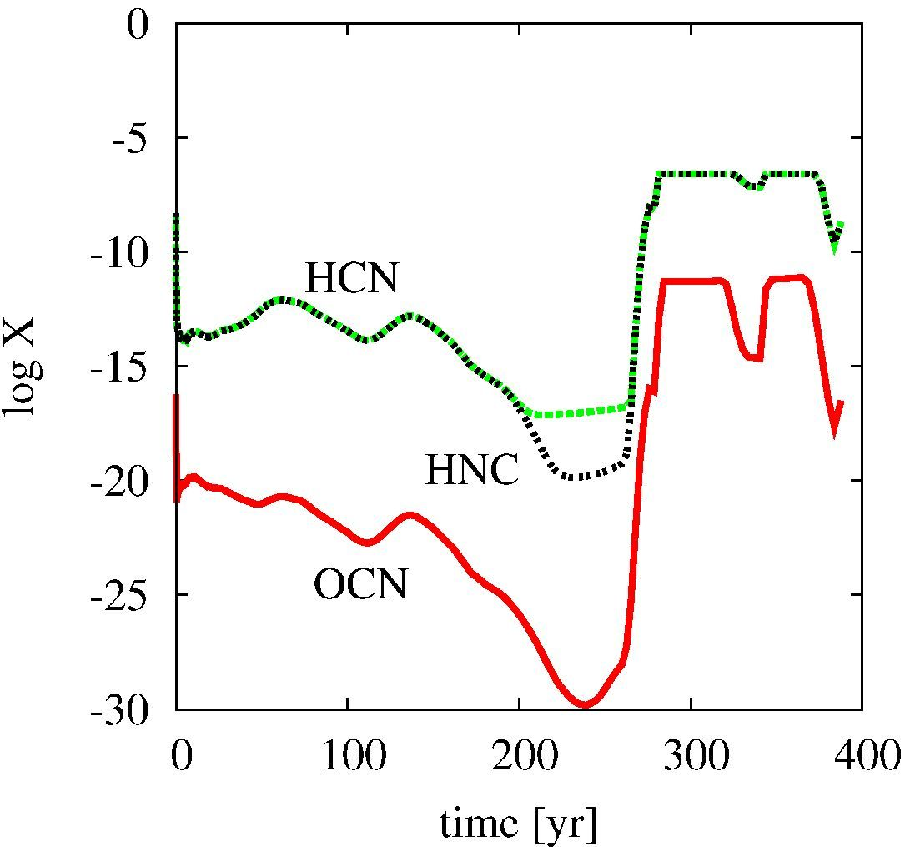}
\hspace{0.2in}
\includegraphics[width=0.8\columnwidth]{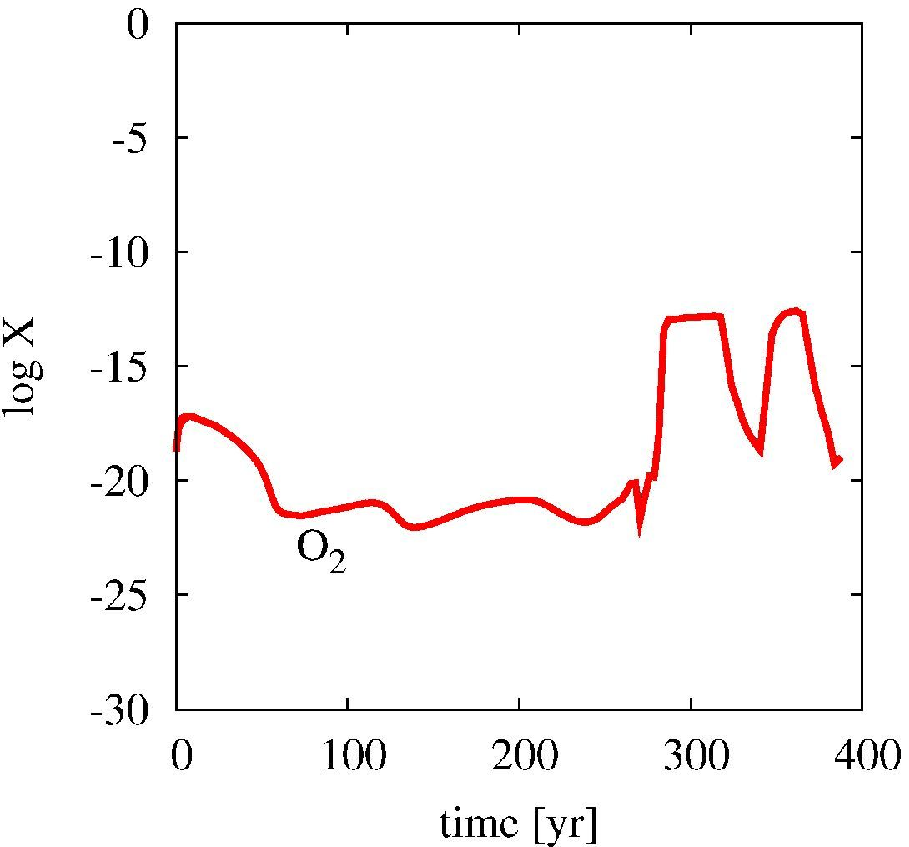}

  \caption{Fractional abundances of species as functions of time for
  the fluid element for which the physical structure is shown in Fig.
  \ref{fig:tracks}.  The relatively cold, quiescent period for the
  first $270\,$yr shows the abundance of most species decreasing due
  to adsorption, or staying constant.}

\label{fig:225chem}
\end{figure*}

Figure \ref{fig:225chem} shows the fractional abundances of 17 species
(CO, SO, NH$_{3}$, H$_{2}$O, H$_{2}$S, OCS, O$_{2}$, HCO, HCO$^{+}$,
HNO, SO$_{2}$, CS, HCS, HCS$^{+}$, HCN, HNC and OCN) as functions of
time during the final cycle for the fluid element for which the
density and temperature are given as functions of time in Fig.
\ref{fig:tracks}.
The shocks induce increases in the gas-phase fractional abundances of
all 17 species.  However, the increase in $X(\rm{CO})$ is too small to
be apparent from the figure, because nearly all CO is in the gas-phase
even in the coldest regions of the disc.

The increases of the fractional abundances of species for which
results are shown in the upper two right hand panels are due almost
entirely to thermal desorption from the surfaces of dust grains.  We
define $N_{\rm{g}}(i)$ to be the average number of molecules of the
$i$th species on a grain. For each species for which results are given
in those panels, $N_{\rm{g}}(i) \eta + X(i)$ is nearly constant.  This
is due to adsorption and desorption being the only processes
significantly affecting the gas-phase fractional abundances of these
species, and the fractional abundance of each of them depends almost
entirely on the temperature only.

The fractional abundances of species for which results are shown in
the other panels are affected significantly by gas-phase reactions, as
well as desorption and adsorption. The higher temperatures in the
shocks allow some reactions with activation energies and some
endothermic reactions to occur at significant rates. Also species
which undergo reactions that are exothermic and have small or no
barriers are desorbed into the gas-phase in shocked regions.

HCS is one species having a fractional abundance that is affected by
gas phase reactions.  It is formed by the reaction of S and CH$_{2}$.
CS is formed by reactions of C with SO, and S with CN in almost equal
proportions. OCN is formed by the reaction of CN and O$_{2}$. HNO
forms by the reaction of NH$_{2}$ and O, and HCO forms mainly via
NH$_{3}^{+}$ reacting with H$_{2}$CO. The reaction of O with OH
contributes to O$_2$ production throughout much of the disc; O$_2$ is
also formed in the shocked regions by the reactions of He$^+$ with
CO$_2$ and SO$_2$.

\subsection{The entire disc}
\label{sec:interpolate}

A major aim of the work reported here is the identification of the
general features that should appear in images of gravitationally
unstable protoplanetary discs obtained in molecular line
emissions. Thus, we present model column density maps for a number of
molecular species.

The final spatial positions of the fluid elements constitute an
irregular grid for which the model fractional abundances are known.
The \textsc{qhull} and \textsc{qgrid3} interpolation routines
contained within the IDL 5.5 libraries were used to obtain the
fractional abundances at the points of a three-dimensional regular
Cartesian grid. The mass density at each of those grid points was
calculated from the mass density distribution at the final time given
by the full hydrodynamic results.  We used the mass density
distribution given by the full fluid results because the grid for
which it was calculated is much more uniform than the grid for which
the chemical results are available. Thus, for regions of low mass
density, we were able to obtain more accurate interpolated results for
the mass density than for the fractional abundances.

\begin{figure*}
 \centering
\includegraphics[width=0.28\textwidth]{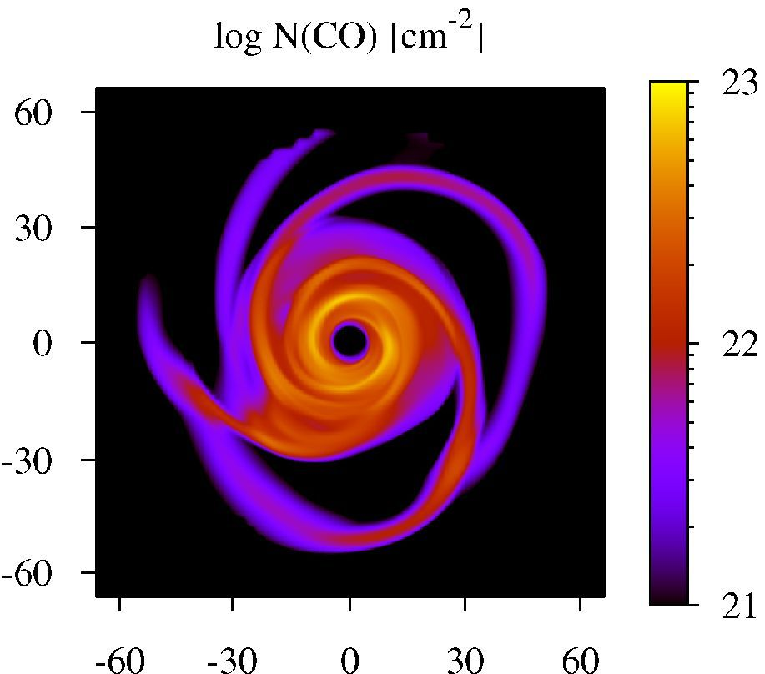}
\hspace{0.2in}
\includegraphics[width=0.28\textwidth]{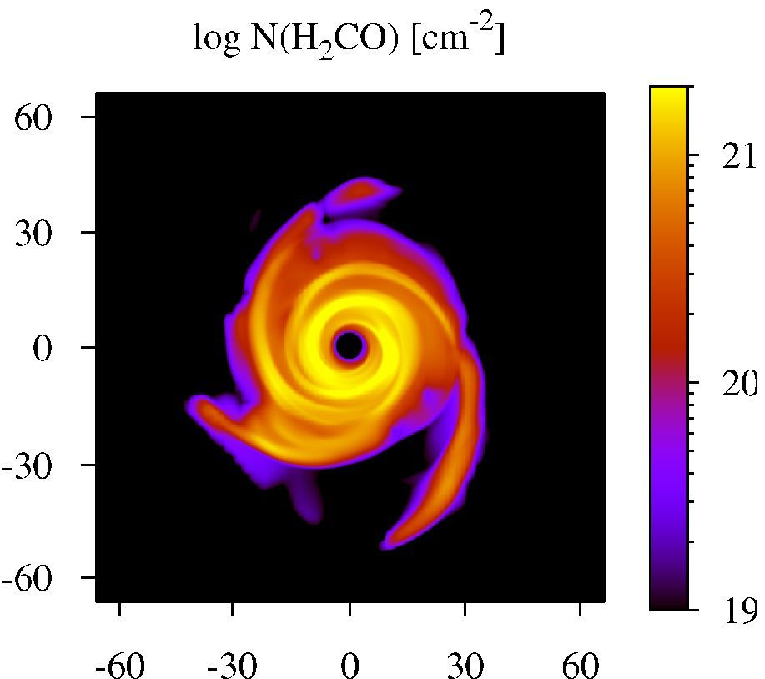}
\hspace{0.2in}
\includegraphics[width=0.28\textwidth]{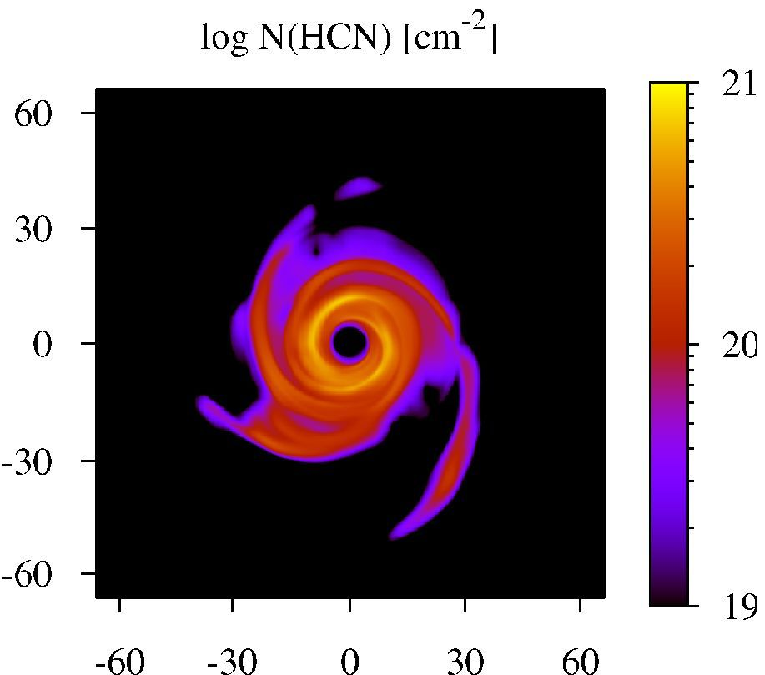}

\vspace{0.1in}

\includegraphics[width=0.28\textwidth]{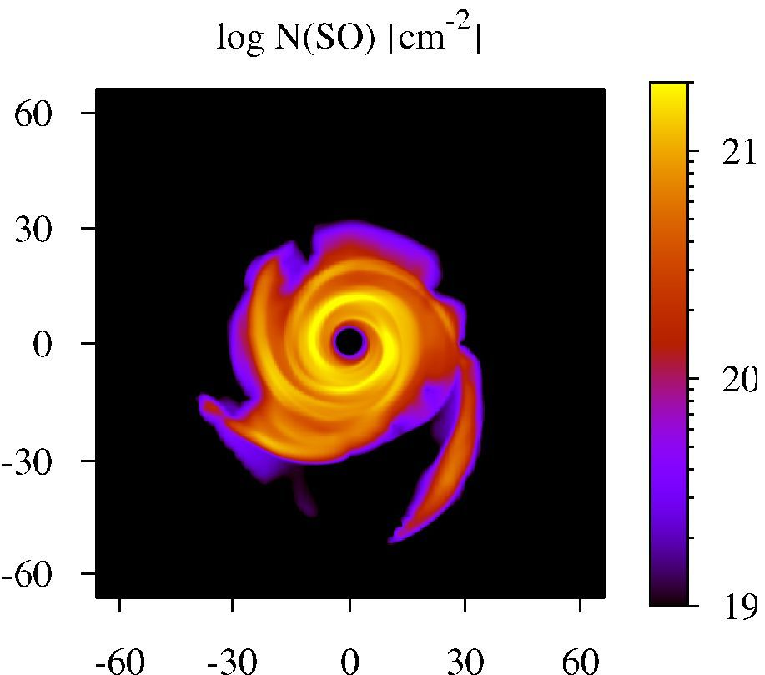}
\hspace{0.2in}
\includegraphics[width=0.28\textwidth]{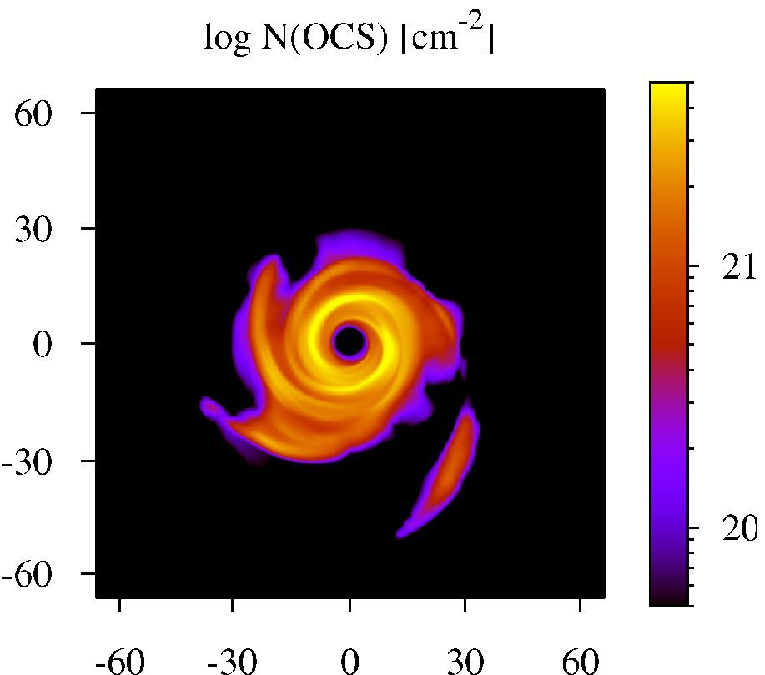}
\hspace{0.2in}
\includegraphics[width=0.28\textwidth]{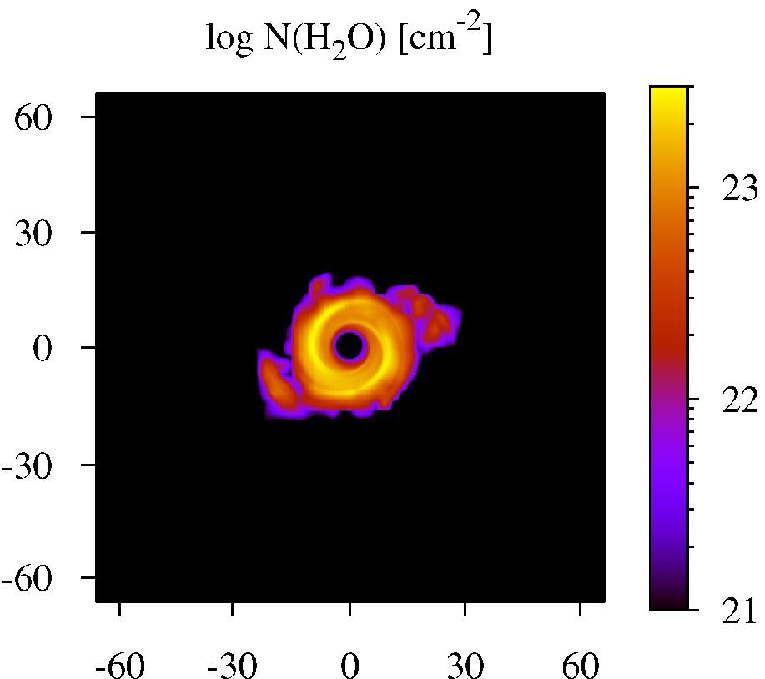}

\vspace{0.1in}

\includegraphics[width=0.28\textwidth]{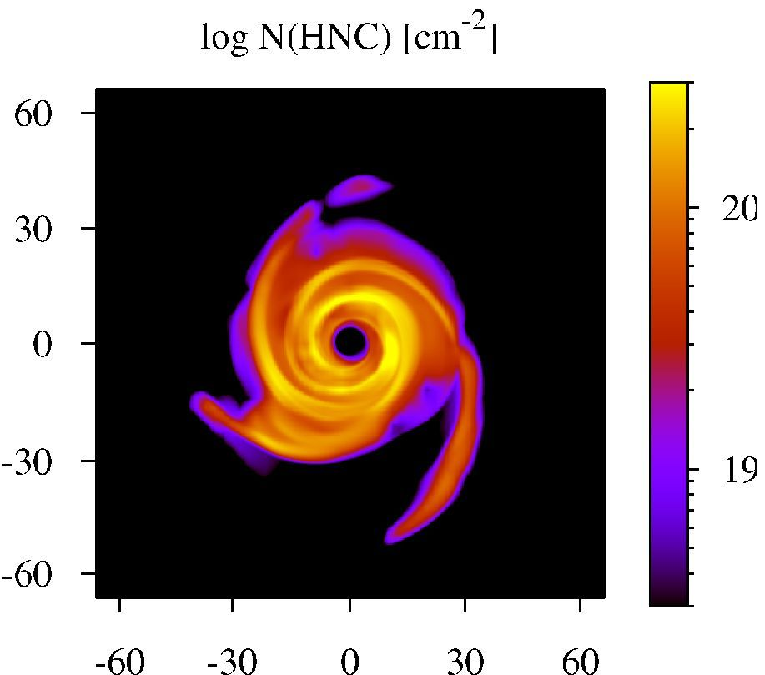}
\hspace{0.2in}
\includegraphics[width=0.28\textwidth]{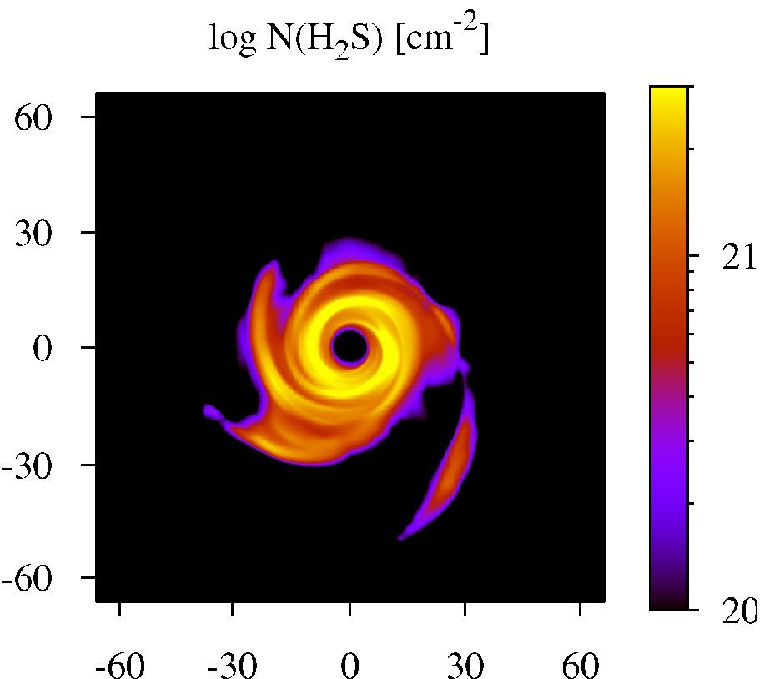}
\hspace{0.2in}
\includegraphics[width=0.28\textwidth]{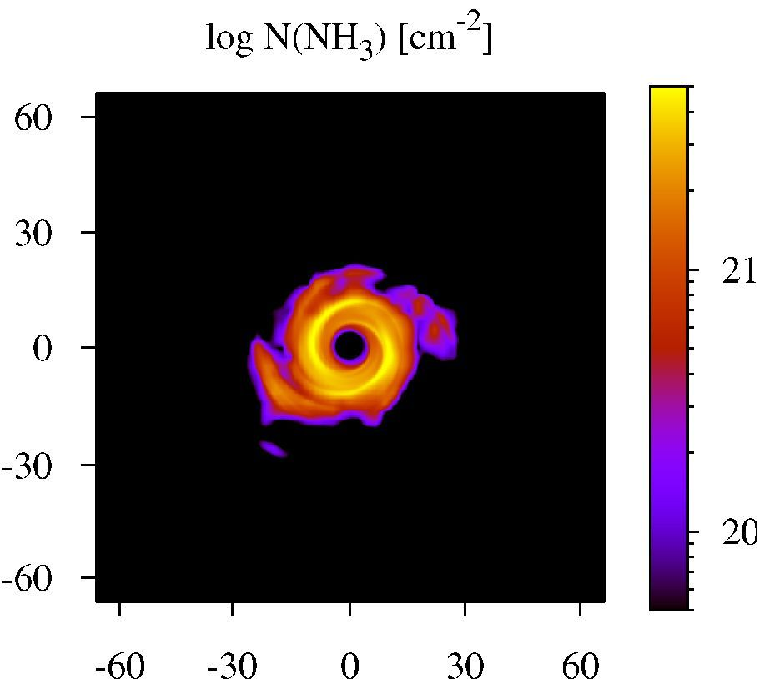}

  \caption{gas-phase column densities of molecules having gas-phase
fractional abundances determined primarily by desorption and
adsorption, at $t=388\,$yr.  Distances from the disc centre are in
au.}
  \label{fig:COetc}
\end{figure*}

\begin{figure*}
 \centering

\vspace{0.2in}
\includegraphics[width=0.28\textwidth]{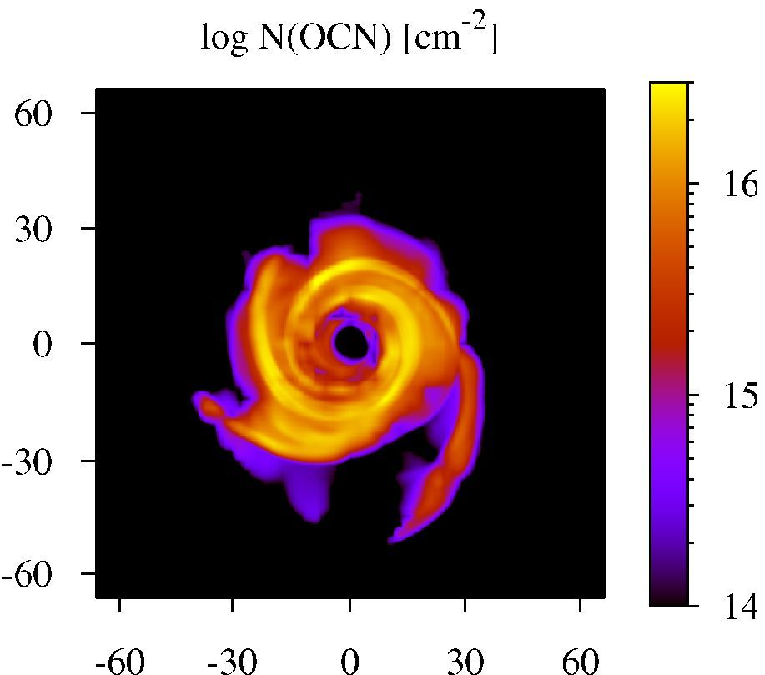}
\hspace{0.2in}
\includegraphics[width=0.28\textwidth]{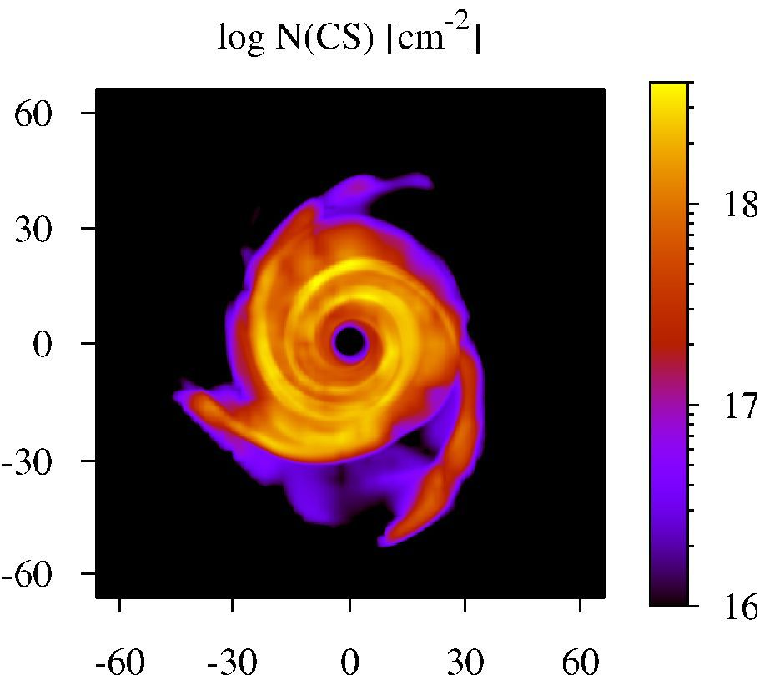}
\hspace{0.2in}
\includegraphics[width=0.28\textwidth]{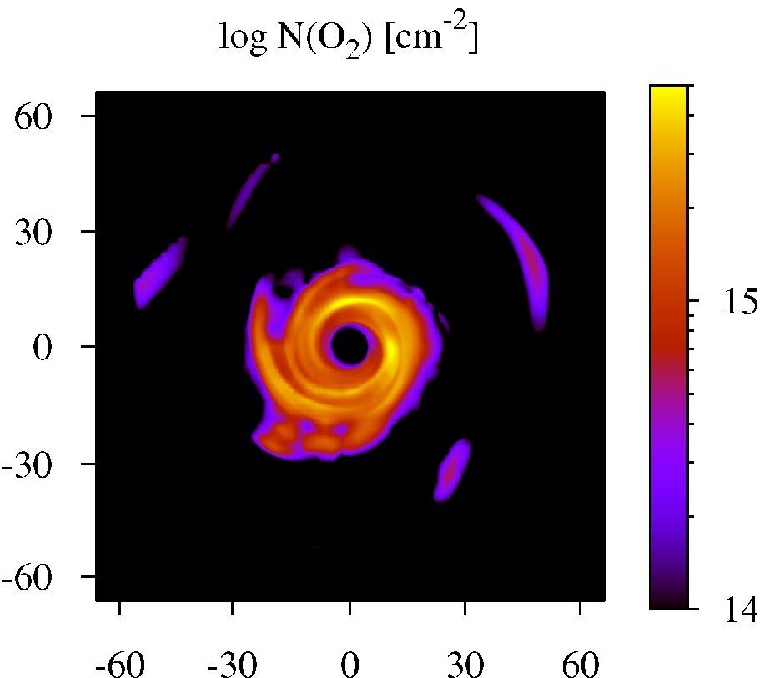}

\vspace{0.1in}

\includegraphics[width=0.28\textwidth]{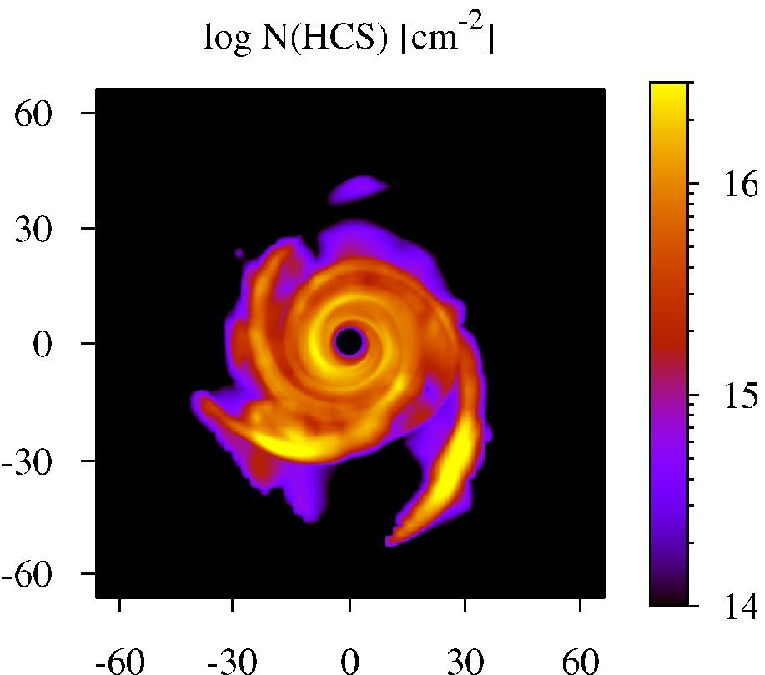}
\hspace{0.2in}
\includegraphics[width=0.28\textwidth]{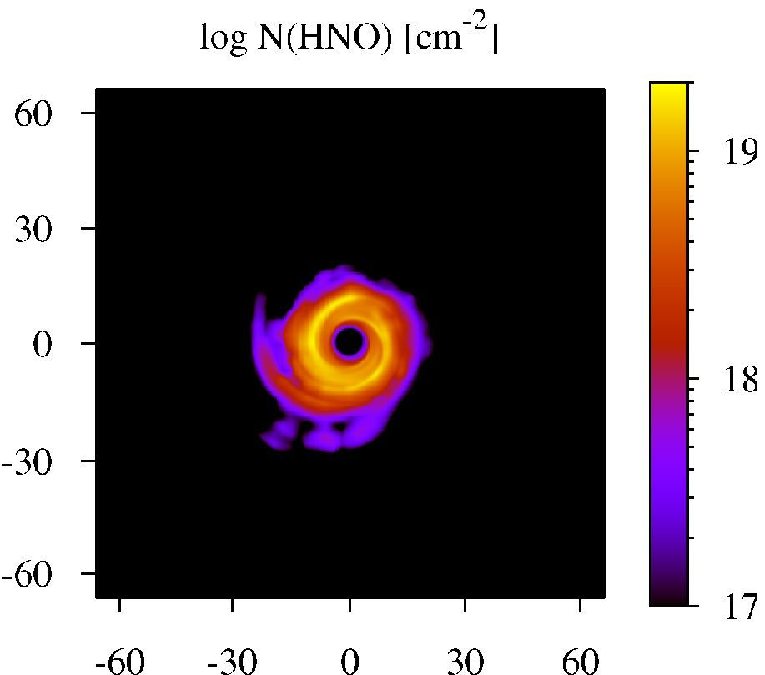}
\hspace{0.2in}
\includegraphics[width=0.28\textwidth]{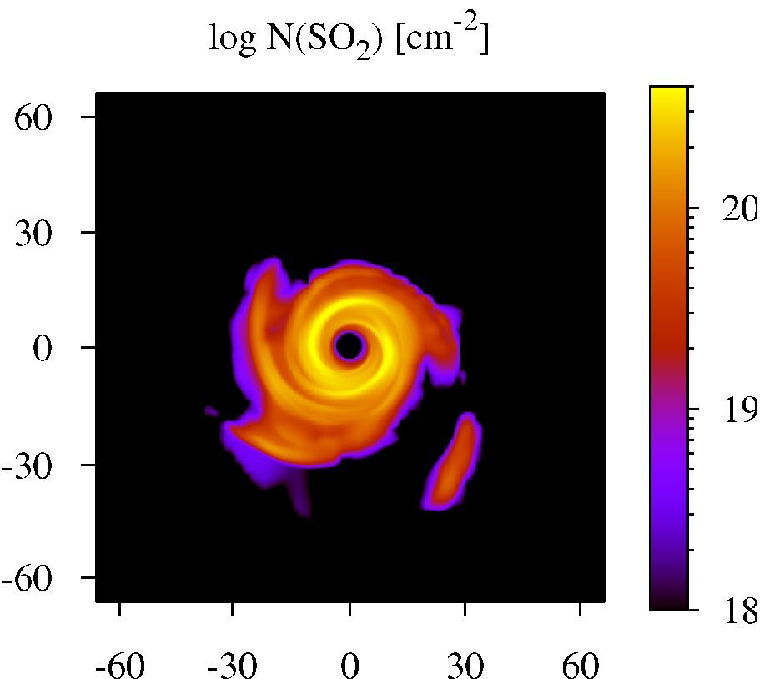}

\vspace{0.1in}

\includegraphics[width=0.28\textwidth]{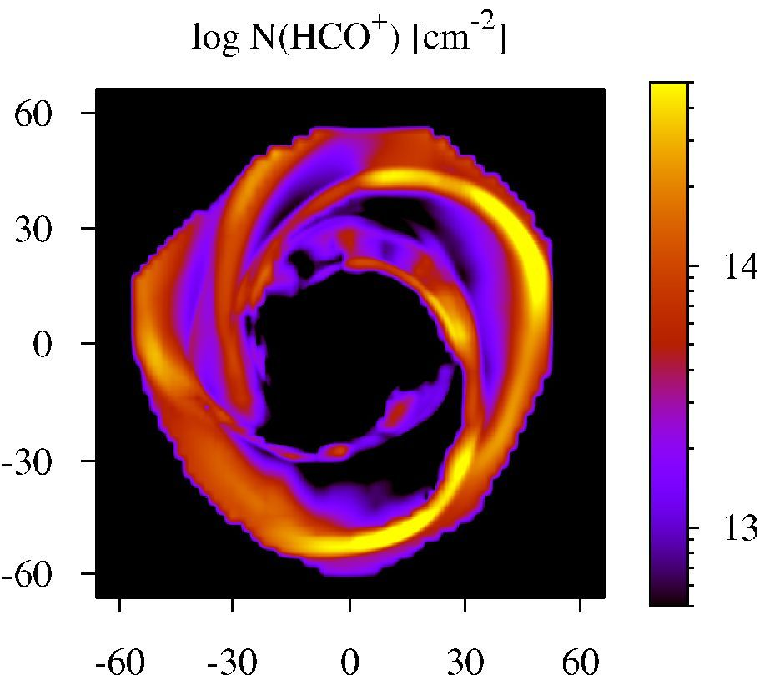}
\hspace{0.2in}
\includegraphics[width=0.28\textwidth]{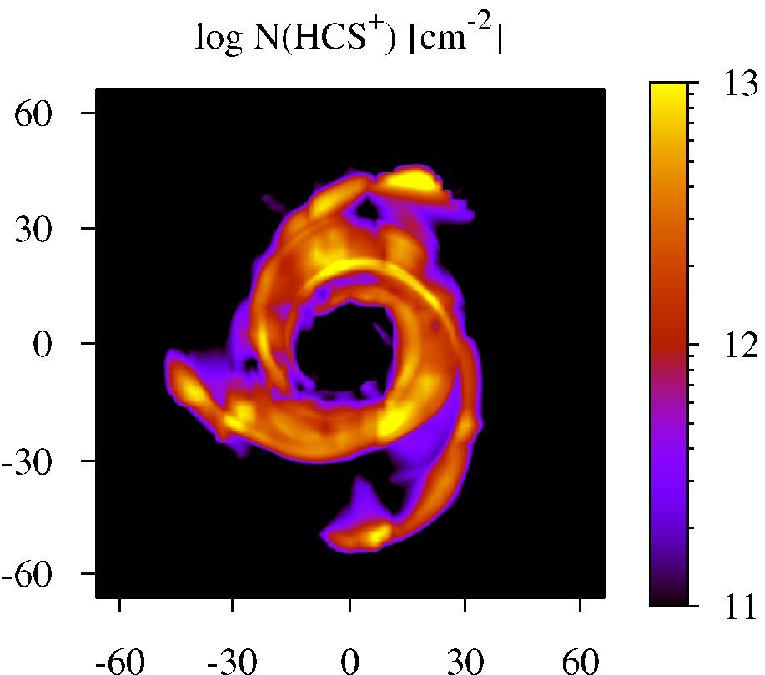}

  \caption{Column densities in the disc for several gas-phase
  molecules whose total abundance varies with time, at $t=388\,$yr.
  Distances from the grid centre are in au.}
  \label{fig:OCNetc}
\end{figure*}

The column density $N(i,x,y)$ of the $i$th species on a line of sight
perpendicular to the disc is
\begin{equation}
N(i,x,y) = \int n X(i,x,y,z) \; \rm{d}z. 
\end{equation} 
We assumed that $n$ and all $X(i,x,y,z)$s are even functions of $z$
when performing the integrals.  Figures \ref{fig:COetc} and
\ref{fig:OCNetc} show the column densities of selected species.
For each species for which results are given in Fig. \ref{fig:COetc},
$N_{\rm{g}}(i) \eta + X(i)$ is nearly constant.  This is due to
adsorption and desorption being the only processes significantly
affecting the gas-phase fractional abundances of these species. Thus,
the structure seen in each of these column densities is due to the
variations in temperature and density and is not influenced by
gas-phase chemistry.

$N_{\rm{g}}(i) \eta + X(i)$ is not constant for each of the species
for which Fig. \ref{fig:OCNetc} shows a gas-phase column density map.
Gas-phase chemistry significantly affects the gas-phase fractional
abundance of each of these species.  We have already given some
explanation for the evolution of most of the species for which results
are displayed in Fig. \ref{fig:OCNetc} in Section
\ref{sec:individual}.  The HCO$^{+}$ is anticorrelated with the
H$_{2}$O and other species with which it charge exchanges.  HCS$^{+}$
is also removed by charge exchange with H$_{2}$O, but its distribution
differs somewhat from that of HCO$^{+}$.  This is due to CS being
abundant in the gas-phase only in regions in which it is efficiently
desorbed and CO being abundant everywhere in the disc.  Reactions of
H$_{3}^{+}$ with CS and CO produce these ions.

\subsection{The effect of three-body reactions}

The chemistry in each of several fluid elements was calculated both
with the three-body reactions included and with them excluded.  As
expected, the largest differences occur in the denser inner regions of
the disc and spiral arms. Even there, the effects were insignificant
for all species except some with low fractional abundances ($X(i) <
10^{-10}$).

A reduction of the fractional abundance of O$_{2}$ in one element from
approximately $10^{-14}$ to $10^{-21}$ in a shocked region and an
enhancement in the fractional abundance of NH from approximately
$10^{-17}$ to $10^{-14}$ were amongst the most pronounced effects of
including the three-body reactions.  Species with fractional
abundances higher than approximately $10^{-10}$ were not noticeably
affected by the addition of the three-body reactions.  The results
that we present were obtained with the three-body reactions included.

\subsection{Comparison with other models} 

The three-dimensional nature of our chemical model and the
qualitatively different dynamics of the disc make a direct comparison
of our results with the previously existing chemical results for
axisymmetric models of lower mass discs difficult.  However, we give a
rough comparison between the maximum fractional abundances of species
given by our model and some obtained by others who have investigated
chemistry for disc models.  The results taken from the other papers
are somewhat inaccurate because they were obtained through the
examination of figures from which precise information was hard to
infer.  Table \ref{tab:compare} shows the maximum fractional
abundances recorded at the end of our simulation and the corresponding
values from \cite{walsh_chemical_2010} and
\cite{ilgner_transport_2004} (hereafter W2010 and I2004 respectively).
\begin{table}
\caption{Maximum abundances reported in this work, W2010 and I2004.  A
dash signifies no data for that species were available in the original
paper.}
\label{tab:compare}
\centering
\begin{tabular}{cccc}
\hline
  Species    & \multicolumn{3}{c}{Maximum $\log X(i)$}   \\
     $i$        & This work & W2010 & I2004 \\
\hline
HCO$^+$      & -10.8  & -6       & -14           \\
HCN          & -6.4   & -6.5     & -             \\  
CN           & -7.8   & -7.5     & -             \\
CS           & -8.0   & -8       & -10           \\
C$_{2}$H     & -10.5  & -7       & -             \\
H$_{2}$CO    & -5.7   & -9       & -             \\
N$_{2}$H$^+$ & -19.6  & -11      & -             \\
H$_{2}$O     & -3.7   & -4       & -             \\
CO$_{2}$     & -4.5   & -4.5     & -             \\
OH           & -12.8  & -4       & -14           \\
S            & -9.9   & -        & -15           \\
SO           & -5.8   & -        & -10           \\
SO$_{2}$     & -6.5   & -        & -7            \\
HCS$^+$      & -12.5  & -        & -21           \\
NH$_{3}$     & -5.5   & -        & -5            \\ 
\hline
\end{tabular}
\end{table}
Our results are comparable to those of the other models for NH$_3$,
SO$_{2}$, H$_{2}$O, CO$_2$, CN, CS and HCN.  We obtain a
higher fractional abundance of HCO$^+$ than I2004, but less than
reported in W2010.  Our fractional abundances of C$_2$H and
N$_2$H$^{+}$ are lower than those of W2010. However, we report a much
higher fractional abundance of H$_2$CO.

Some of the discrepancies between the results of W2010 and
I2004 arise because W2010 considered a much more extensive fraction of
the disc than I2004 who studied only the material at radii less than
$10\,$au. We studied material at radii between $10$ and $60\,$au,
where spiral structure and shocks are important in our dynamical
model. Thus, in further discussion of the differences between our
results and those of others, we focus on a comparison of our results
and those of W2010.

Some of these differences are due to the assumptions that we made
concerning the initial chemical conditions. As mentioned earlier, in
our model a number of species have gas-phase abundances that depend
only on the assumed initial conditions and the balance between
desorption and absorption. H$_2$CO in particular is such a
species.  So its high abundance in our model compared to that found by
W2010 is due to our assumption of a higher H$_2$CO abundance and
neglect of species as massive as CH$_3$OH. Similarly, our assumption
that no nitrogen is initially in N$_2$ leads to a lower abundance of
N$_2$H$^+$ than W2010 found.

The high values of the HCO$^+$ and C$_2$H given for W2010
occur in regions with densities that are about four orders of
magnitude lower than any that we consider; these regions have
correspondingly lower total column densities than the much denser
regions do and in the W2010 model, the chemistry in them is affected
by the diffusion that W2010 assume to occur, but which we do not.  The
mid-plane value obtained by W2010 for the fractional abundance of
HCO$^+$ for radii of $10$ to $60\,$au is comparable to the maximum
value that we report for it; this is to be expected because in both
models most of the CO is in the gas phase, and HCO$^+$ is formed by a
simple ionisation triggered reaction sequence. The mid-plane value
obtained by W2010 for the fractional abundance of C$_2$H for radii of
$10$ to $60\,$au is less than the maximum value that we report for it;
this is due to the desorption, caused by shock induced heating,
occurring near the mid-plane in our models.

\section[Conclusions]{Conclusions \& Future Work}
\label{sec:conclusions}

We have constructed a chemical model of a $0.39\,$M$_{\odot}$
protoplanetary disc, surrounding a solar-mass star, representative of
a Class 0 or early Class I system.  In the disc, gravitational
instabilities cause spiral waves.  The shocks associated with these
waves induce the desorption of various chemical species from the
surface of dust grains and increase their gas-phase abundances. Though
the gas-phase fractional abundances of some of the species are not
significantly affected by gas-phase chemistry, some of the desorbed
species are reactive. In some cases, the elevated temperatures in the
shocked regions enhance the reactivity.

Because most of the mass is concentrated in spiral structures, all of
the gas-phase molecular column densities show spiral
structures. However, the structures are limited in extent in
some species, e. g.\ H$_2$O, which possess the highest binding energies
to grains, therefore higher temperatures are needed for desorption to
occur.  This also limits the extent of structure in species which form
in the gas-phase from these tightly bound species.  Consequently,
maps of the emissions of a number of species will reveal where shocks
of differing strengths are and, thus, serve as diagnostics of the disc
dynamics.

We find that three-body reactions have little effect on the chemistry
of species with fractional abundances above $10^{-10}$.  Though they
are most important in the hotter, densest regions of the disc, they do
not alter the overall chemistry of the disc significantly.

A direct comparison of our results with those of other models is not
straightforward, due largely to the very different nature of the disc
dynamics that we have used.  Some differences between the peak gas
phase fractional abundances that we obtained and those reported in
other papers are due to the assumptions that we have made about
chemical initial conditions.  However, the major differences in the
chemistry away from the outer boundaries of the disc arise from the
role that some other modellers assume that diffusion plays in
enhancing the richness of the chemistry in the disc interior
\citep[e.g.][]{ilgner_transport_2004,heinzeller_chemical_2011}; in
many of these models, the chemistry at the midplane is not very rich
over a very large fraction of the disc. In contrast, our model gives
rise to substantial gas-phase abundances near the midplane,
even though we have not assumed that efficient microscopic
mixing is a consequence of large-scale turbulence. Though
inclusion of such mixing would be difficult in an approach in which
individual fluid elements are followed, we have not neglected that
type of mixing for that reason. Rather, we have chosen to focus on
how gravitational instability generates shocks which induce a rich
gas-phase chemical composition in the disc interior, even near the
midplane (see Appendix \ref{sec:persp}).

ALMA will revolutionise observational studies of discs.  If used to
observe a disc in the Taurus-Auriga cloud complex, ALMA, with its
$5\,$milliarcsec resolution, will allow the mapping of features on
scales of about $1\,$au, which is smaller than the widths of the
prominent spiral structures of our model.  We plan to use our results
in radiative transfer calculations to obtain reliable estimates of the
detectability of structure in different species.

\section*{Acknowledgments}

We are grateful to Pierre Lesaffre for a thorough and helpful
referee's report.  We have benefitted from the use of data in the
online database KIDA (KInetic Database for
Astrochemistry\footnote{\url{http://kida.obs.u-bordeaux1.fr}}).
Resources supporting the hydrodynamics simulation were provided by the
NASA High-End Computing (HEC) Program through the NASA Advanced
Supercomputing (NAS) Division at Ames Research Center.  JDI gratefully
acknowledges a studentship from the Science and Technology Facilities
Council of the United Kingdom (STFC).  ACB's contribution was
supported in part by the National Science Foundation under Grant
No.~PHY05-51164 and in part under contract with the California
Institute of Technology (Caltech) funded by NASA through the Sagan
Fellowship Program.  PC's and TWH's work on star and planet formation
is supported by a STFC rolling grant.  RHD was supported in part by
NASA Origins of Solar Systems grant NNX08AK36G.  TWH appreciates
funding provided by the Indiana University Institute for Advanced
Study.


\bibliographystyle{mn2e} 

\appendix

\section[Other perspectives]{Other perspectives}
\label{sec:persp}

\begin{figure*}
 \centering
\includegraphics[width=0.9\columnwidth]{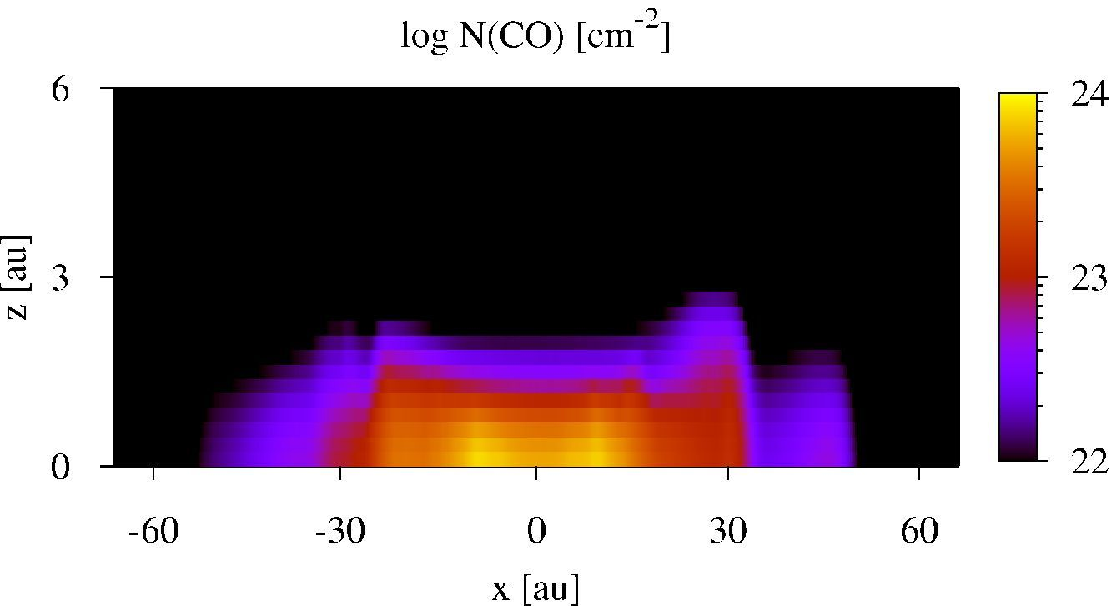}
\hspace{0.2in}
\includegraphics[width=0.9\columnwidth]{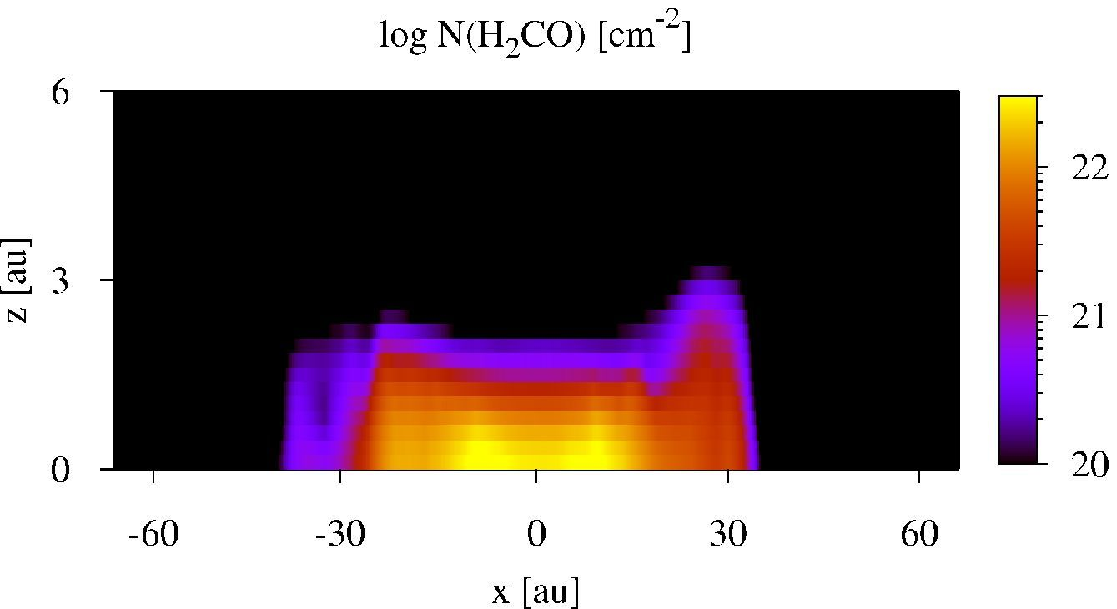}

\vspace{0.2in}

\includegraphics[width=0.9\columnwidth]{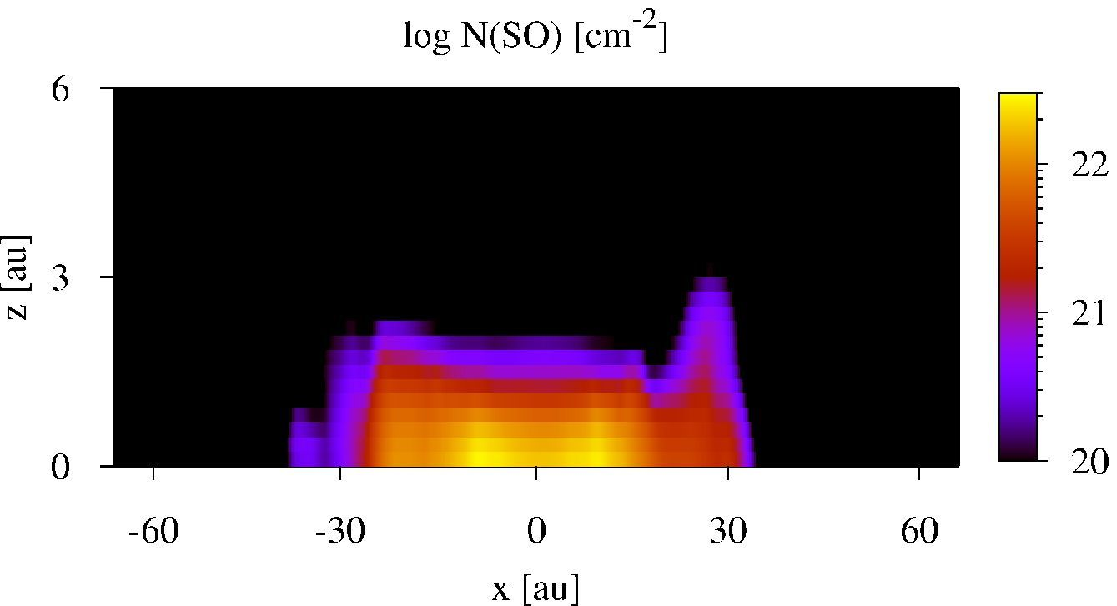}
\hspace{0.2in}
\includegraphics[width=0.9\columnwidth]{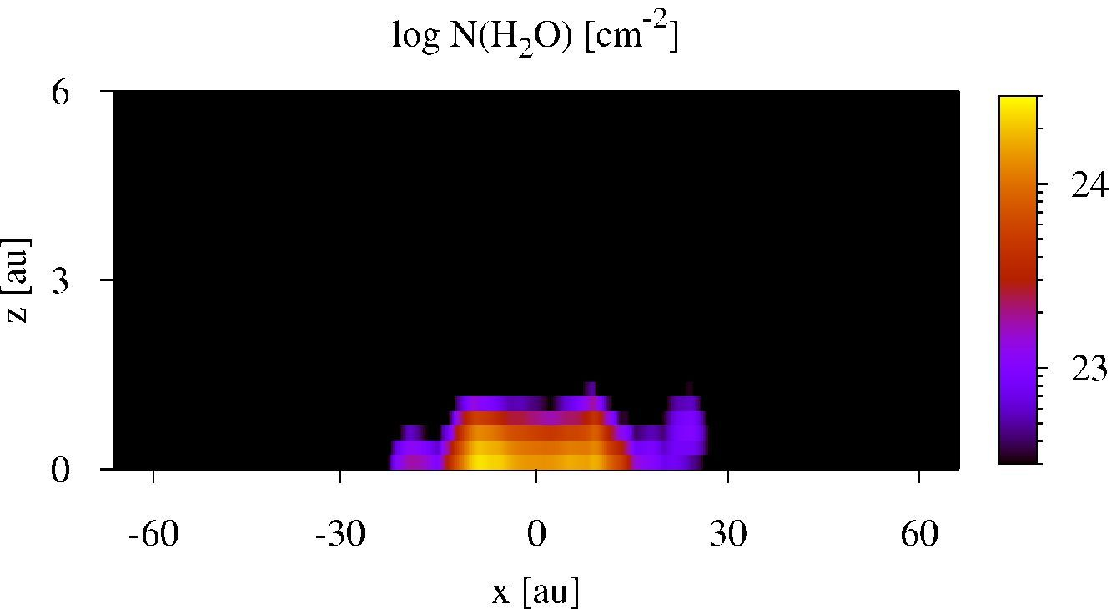}
  \caption{Column densities of the disc as viewed along the $y$-axis
  of Fig. \ref{fig:Nnden}, towards $y=0$, at $t=388\,$yr.}
   \label{fig:sides}
\end{figure*}

Figure \ref{fig:sides} shows the column densities of selected species
in the disc as viewed along the $y$-axis, toward $y=0$, in Fig.
\ref{fig:Nnden}.  The majority of the edge-on maps show little
structure (as with CO and H$_{2}$O), though some, such as the
H$_{2}$CO and SO maps, show slightly higher column densities on the
right side due to the presence of a large spiral arm.

\begin{figure*}
\centering
\includegraphics[width=0.9\columnwidth]{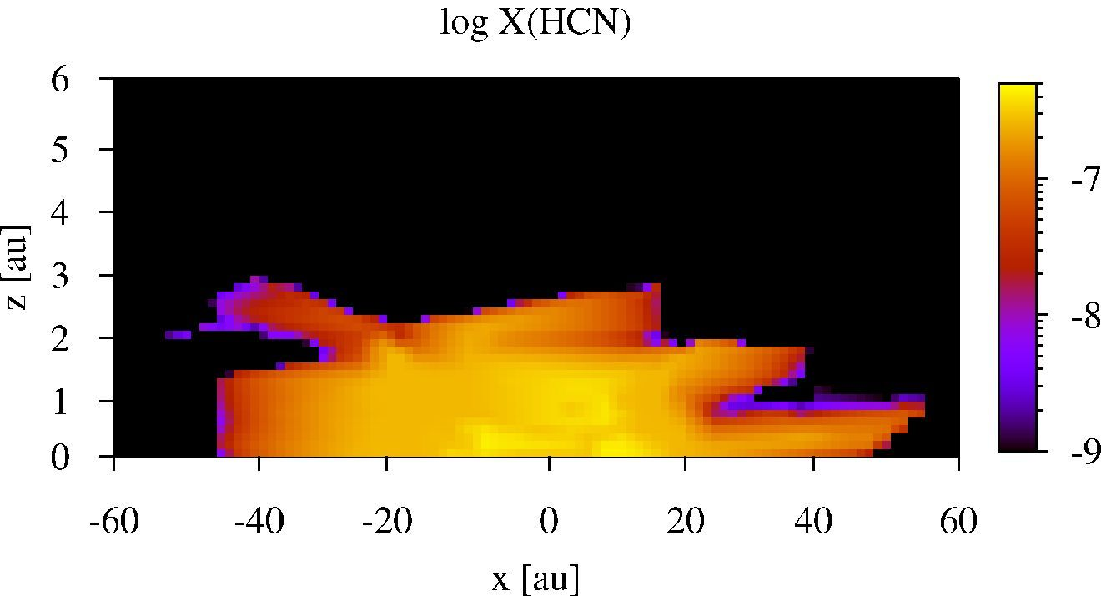}
\hspace{0.2in}
\includegraphics[width=0.9\columnwidth]{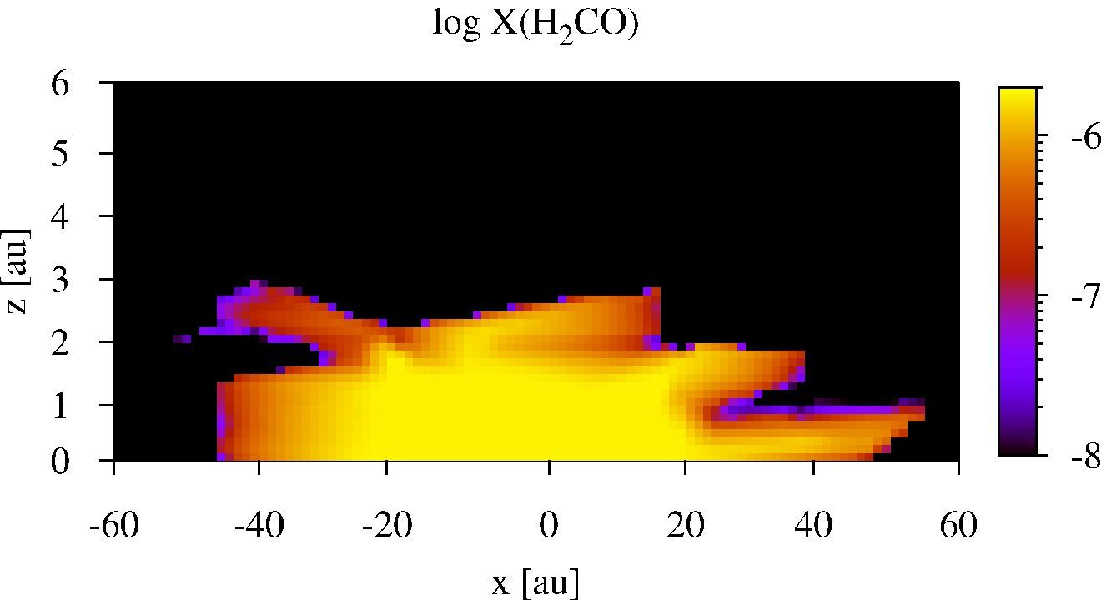}

\vspace{0.2in}

\includegraphics[width=0.9\columnwidth]{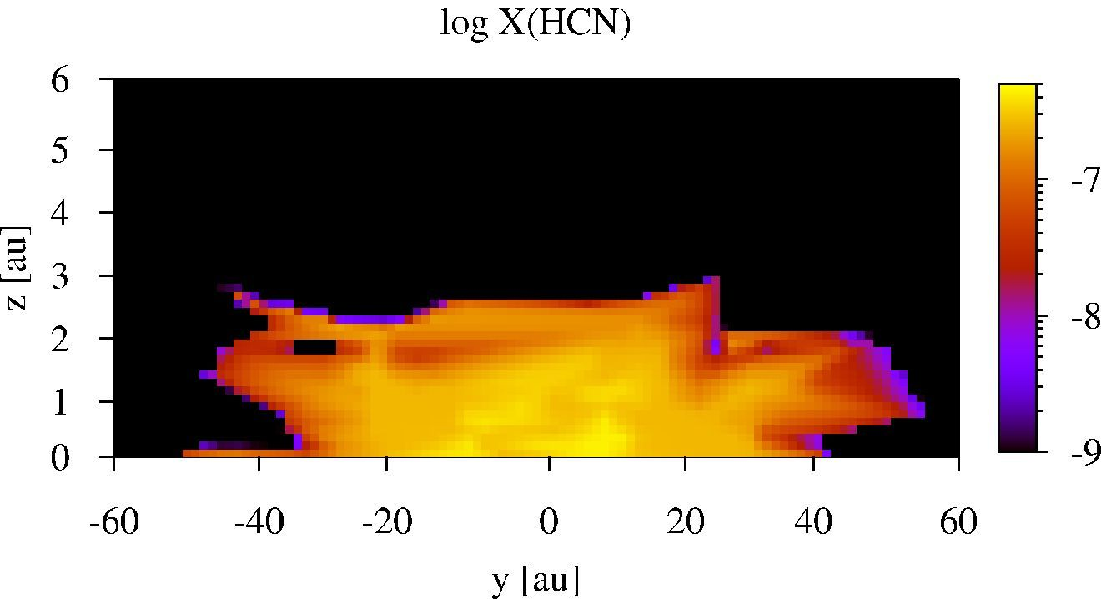}
\hspace{0.2in}
\includegraphics[width=0.9\columnwidth]{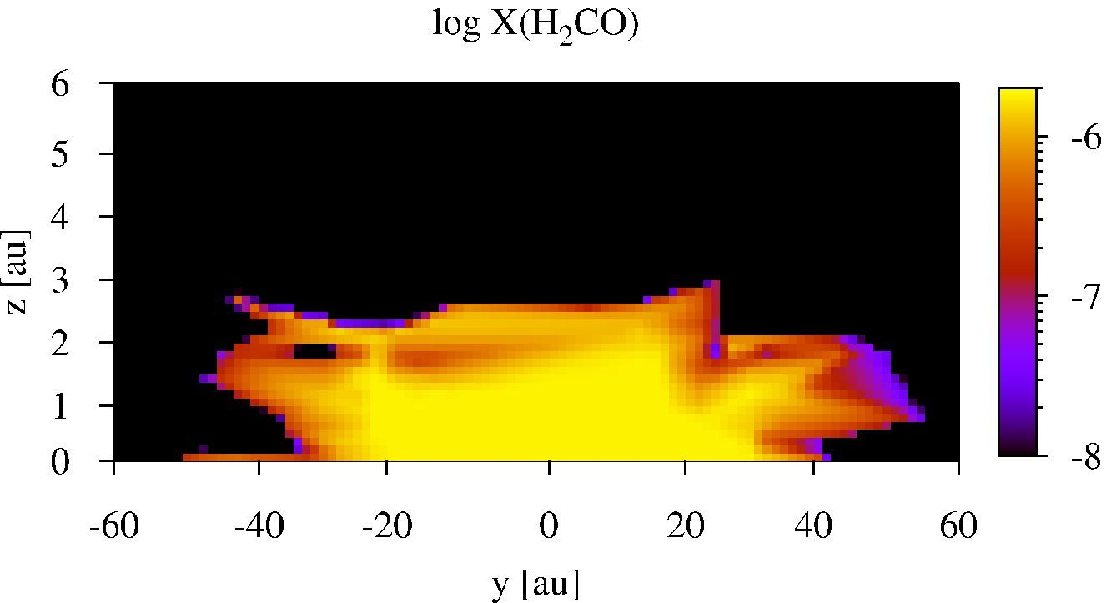}

  \caption{Slices of the fractional abundance of disc interior,
  oriented as in Figure \ref{fig:tnslice}, interpolated from the
  positions of the fluid elements at $t=388\,$yr.}
  \label{fig:slices}
\end{figure*}

Figure \ref{fig:slices} shows slices of the fractional abundances in
the interior of the disc for HCN and H$_{2}$CO.  These show little
structure in the disc interior when compared with other models of
axisymmetric discs.  These shock-desorbed molecules show little
structure in the disc interior compared with what is seen in
axisymetric $\alpha$-disc models.

\section[Three-body reactions]{Three-body reactions}
\label{sec:3br}

Table \ref{tab:3br} lists the three-body reactions included in the
chemical network and taken from \cite{willacy_gas_1998} and the more
recent UMIST Database for Astrochemistry,
UDfA\footnote{\url{http://www.udfa.net}}.  The subset includes only
the reactions possessing activation energies of less than $10^{4}\,$K.

\begin{table*}
\caption{Three-body reactions included in the chemical network.  In
all but one of the reactions the third body is assumed to be molecular
hydrogen. References are W: \citet{willacy_gas_1998} and U:
\citet{woodall_umist_2007}.}
\label{tab:3br}
\centering
\begin{tabular}{ccccccccccc}
\hline
$R_{1}$ & $R_{2}$ & $R_{3}$ & $\rightarrow$ & $P_{1}$ & $P_{2}$ & $\alpha$ & $\beta$ & $\gamma$ & Reference \\
\hline

 CH$_3$  &   H   & H$_{2}$  & $\rightarrow$ &    CH$_4$   &  H$_{2}$       &       9.62(-31) & -1.80 & 3231.0 & W	\\
 N    &   O   & H$_{2}$  & $\rightarrow$ &   NO  &    H$_{2}$    &          3.86(-34) &  0.00 & 2573.0 & W \\
 O    &   O   & H$_{2}$  & $\rightarrow$ &   O$_2$  &    H$_{2}$       &       5.25(-35) &  0.00 &  902.0 & W \\
 O   &    H   & H$_{2}$  & $\rightarrow$ &   OH  &    H$_{2}$       &       6.23(-34) &  0.00 & 1960.0 & W \\
 OH   &   H   & H$_{2}$  & $\rightarrow$ &   H$_{2}$O &    H$_{2}$       &       6.86(-31) & -2.00 &    0.0 & W \\
 C    &   O   & H$_{2}$  & $\rightarrow$ &   CO  &    H$_{2}$    &          2.14(-29) & -3.08 & 2441.0 & W \\
 CO   &   O   & H$_{2}$  & $\rightarrow$ &   CO$_2$ &    H$_{2}$       &       9.56(-34) &  0.00 & 1060.0 & W \\
 H    &   CO  & H$_{2}$  & $\rightarrow$ &   HCO &    H$_{2}$    &          5.30(-34) &  0.00 & 370.0 & W \\
 SO   &   O   & H$_{2}$  & $\rightarrow$ &   SO$_2$ &    H$_{2}$       &       1.86(-31) & -0.50 &    0.0 & W \\  
 H   &    H   & H$_{2}$ & $\rightarrow$ &    H$_2$  &    H$_2$   &          8.72(-33) & -0.60 &   0.0 & W \\ 
 H   &    H   & H  & $\rightarrow$ &   H     &  H$_2$    &          1.83(-31) & -1.00 &   0.0 & W \\ 
 N  &     N   & H$_{2}$  & $\rightarrow$ &    N$_2$  &    H$_{2}$       &       9.53(-34) & -0.50 &   0.0 & W \\
 N    &   H   & H$_{2}$  & $\rightarrow$ &    NH &     H$_{2}$    &          3.09(-32) & -0.50 &  0.0 & W \\
 NH  &    H   & H$_{2}$  & $\rightarrow$ &    NH$_2$  &   H$_{2}$       &       3.18(-33) & -0.50 &   0.0 & W \\
 NH$_2$  &   H   & H$_{2}$  & $\rightarrow$ &   NH$_3$  &   H$_{2}$       &       1.32(-33) &  0.00 & 4366.0 & W \\
 H  &     CH$_3$ & H$_{2}$  & $\rightarrow$ &    CH$_4$  &   H$_{2}$       &       6.16(-29) & -1.80 &  0.0 & U \\
 H &      O   & H$_{2}$  & $\rightarrow$ &    OH   &   H$_{2}$    &          4.33(-32) & -1.00 &  0.0 & U \\
 H &      NH$_2$ & H$_{2}$  & $\rightarrow$ &    NH$_3$ &    H$_{2}$       &       6.06(-30) &  0.00 &  0.0 & U \\
 H   &    OH  & H$_{2}$  & $\rightarrow$ &   H$_2$O  &   H$_{2}$    &          1.51(-31) & -2.65 &  1.9 & U \\
 H  &     CO  & H$_{2}$   & $\rightarrow$ &    HCO  &   H$_{2}$    &          5.30(-34) &  0.00 & 370.0 & U \\
 H    &   HCO & H$_{2}$  & $\rightarrow$ &    H$_{2}$CO &    H$_{2}$      &       3.16(-30) & -2.57 &  215.0 & U \\
 H    &   NO  & H$_{2}$  & $\rightarrow$ &   HNO  &   H$_{2}$    &          1.33(-31) & -1.32 & 370.0 & U \\
 H$_2$   &   C   & H$_{2}$  & $\rightarrow$ &    CH$_2$ &    H$_{2}$       &       7.00(-32) &  0.00 &   0.0 & U \\
 H$_2$  &    CH  & H$_{2}$  & $\rightarrow$ &    CH$_3$  &   H$_{2}$       &       3.79(-30) & -1.84 & 65.4 & U \\
 H$_2$  &    N   & H$_{2}$  & $\rightarrow$ &    NH$_2$  &   H$_{2}$       &       1.00(-26) &  0.00 &  0.0 & U \\
 C   &    N   & H$_{2}$  & $\rightarrow$ &    CN   &   H$_{2}$    &          9.41(-33) &  0.00 &  0.0 & U \\
 O   &    SO  & H$_{2}$  & $\rightarrow$ &   SO$_2$  &   H$_{2}$    &          9.15(-31) & -1.84 &  0.0 & U \\
 O    &   O   & H$_{2}$  & $\rightarrow$ &   O$_2$   &   H$_{2}$    &          1.98(-34) & -0.48 & -564.0 & U \\
 C    &   O   & H$_{2}$  & $\rightarrow$ &   CO   &   H$_{2}$    &          2.14(-29) & -3.08 & -2114.0 & U \\
 C$^+$  &    O   & H$_{2}$   & $\rightarrow$ &   CO$^+$  &   H$_{2}$       &       2.14(-27) & -3.08 & -2114.0 & U \\
 C   &    O$^+$  & H$_{2}$  & $\rightarrow$ &   CO$^+$  &   H$_{2}$       &       2.14(-27) & -3.08 & -2114.0 & U \\
 C   &    C   & H$_{2}$  & $\rightarrow$ &   C$_2$   &   H$_{2}$    &          1.26(-32) & -0.64 & -5255.0 & U \\

\hline
\end{tabular}
\end{table*}

\bsp

\label{lastpage}

\end{document}